\documentclass[runningheads]{svmult}

\usepackage{makeidx}   
\usepackage{graphicx}  
\usepackage{subeqnar}  
\usepackage{multicol}  
\usepackage{physprbb}  
\makeindex             

\begin{document}
\title*{High energy particles from $\gamma$-ray bursts\footnote
{Based on lectures given at the {\it ICTP Summer School on Astroparticle 
Physics and Cosmology} (ICTP, Trieste Italy, June 2000), and at the
{\it VI Gleb Wataghin School on High Energy Phenomenology} 
(UNICAMP, Campinas Brazil, July 2000).}
}
\toctitle{High energy particles from $\gamma$-ray bursts}
%
%
\titlerunning{$\gamma$-ray bursts}
%
\author{Eli Waxman}
\authorrunning{Eli Waxman}
%
%
\institute{Weizmann Institute of Science,
Rehovot 76100, Israel}

\maketitle              

\begin{abstract}

A review is presented of the fireball model of $\gamma$-ray 
bursts (GRBs), and of the production in GRB fireballs of
high energy protons and neutrinos. Constraints imposed on the model
by recent afterglow observations, which support the association of
GRB and ultra-high energy cosmic-ray (UHECR) sources, are discussed. 
Predictions of the GRB model for UHECR production, which can be
tested with planned large area UHECR detectors and with planned
high energy neutrino telescopes, are reviewed.


\end{abstract}

\section{Introduction and summary}

The widely accepted interpretation of the
phenomenology of $\gamma$-ray bursts (GRBs),
bursts of 0.1 MeV--1 MeV photons lasting for a few seconds
(see \cite{Fishman} for review), is that the
observable effects are due to the dissipation of the kinetic energy
of a relativistically expanding wind, a 
``fireball,'' whose primal cause is not yet known
(see \cite{fireballs1,fireballs2} for reviews). 
The recent detection of ``afterglows,'' 
delayed low energy (X-ray to radio) emission 
of GRBs (see \cite{AG_ex_review} for review),
confirmed the cosmological origin of the bursts,
through the redshift determination of several GRB host-galaxies,
and confirmed standard model predictions of afterglows
that result from the collision of an expanding fireball with
its surrounding medium (see \cite{AG_th_review} for review). 
In this review, the production in GRB fireballs of $\gamma$-rays,
high-energy cosmic-rays and neutrinos is discussed in the 
light of recent GRB and ultra-high-energy cosmic-ray observations. 

The fireball model is described in detail in \S\ref{sec:fireball}.
We do not discuss in this section the issue of GRB progenitors, i.e.
the underlying sources producing the relativistic fireballs. At present,
the two leading progenitor scenarios are collapses of
massive stars \cite{Fryer98,BohdanHN}, and mergers of compact objects
\cite{Goodman86,Bohdan86}. As explained in \S\ref{sec:fireball}, 
the evolution
of the fireball and the emission of $\gamma$-rays and afterglow radiation
(on time scale of a day and longer) are largely independent of the nature
of the progenitor. Thus, although present observations provide stringent 
constraints on the fireball model, the underlying progenitors remain 
unknown (e.g. \cite{Livio00}; see \cite{AG_ex_review,AG_th_review} 
for discussion). 
In \S\ref{sec:afterglow}, constraints imposed on the fireball model
by recent afterglow observations are discussed, 
which are of importance for high energy particle production. 

The association of GRBs and ultra-high energy cosmic-rays
(UHECRs) is discussed in \S\ref{sec:UHECR}. 
Recent afterglow observations strengthen the evidence
for GRB and UHECR association, which is based on two key points
(see \cite{My-rev} for recent review).
First, the constraints imposed on fireball model parameters
by recent observations imply that 
acceleration of protons is possible to energy higher than previously
assumed, $\sim10^{21}$~eV. Second, the
inferred local ($z=0$) GRB energy generation rate of $\gamma$-rays,
$\sim10^{44}{\rm erg/Mpc}^3{\rm yr}$, is remarkably similar to 
the local generation rate of UHECRs implied by cosmic-ray observations.

The GRB model for UHECR production makes unique predictions, that may be
tested with operating and planned large area UHECR detectors 
\cite{HiRes,Auger1,Auger2,TA}. These 
predictions are described in \S\ref{sec:CRpred}. 
In particular, a critical energy 
is predicted to exist, 
$10^{20}{\rm eV}\le \epsilon_c<4\times10^{20}{\rm eV}$,
above which a few sources produce most of the UHECR flux, and the 
observed spectra of these sources is predicted to be 
narrow, $\Delta \epsilon/\epsilon\sim1$: the bright sources
at high energy should be absent in UHECRs of much
lower energy, since particles take longer to arrive the lower their
energy. If the sources predicted by this model are detected 
by planned large area UHECR detectors, this would not only confirm
the GRB model for UHECR production, but will also provide 
constraints on the unknown structure and strength of the inter-galactic
magnetic field.

We note, that the AGASA experiment has recently reported the presence
of one triplet and 3 doublets of UHECR events above $4\times10^{19}{\rm eV}$,
with angular separations (within each group) 
$\le2.5^\circ$, roughly consistent with the measurement error
\cite{agasa-multiplets}. The probability that these multiplets are
chance coincidences (as opposed to being produced by point sources) is
$\sim1\%$. Therefore, this observation favors
the bursting source model, although more data are needed to confirm it.
Testing the predictions of the fireball model for UHECR production
would require an exposure 10 times larger than that of present
experiments. Such increase is expected to be provided by the
HiRes \cite{HiRes} and Auger 
\cite{Auger1,Auger2} detectors, and by the proposed
Telescope Array detector \cite{TA}.

Predictions for the emission
of high energy neutrinos from GRB fireballs are discussed in \S\ref{sec:nus}.
Implications for planned
high energy neutrino telescopes (the ICECUBE extension of AMANDA,
ANTARES, NESTOR; see \cite{Halzen-rev99} for review) are discussed in 
detail in \S\ref{sec:implications}. It is shown that 
the predicted flux of $\ge10^{14}$~eV
neutrinos may be detectable by Cerenkov neutrino telescopes
while the flux above
$10^{19}$~eV may be detectable by large air-shower
detectors \cite{Auger-nus,OWL1,OWL2}. Detection of the predicted neutrino
signal will confirm the GRB fireball model for UHECR production and
may allow to discriminate between different fireball progenitor scenarios.
Moreover, a detection of even a handful of neutrino events correlated 
with GRBs will allow to test for neutrino properties, e.g. flavor
oscillation and coupling to gravity, with accuracy many orders of magnitude
better than currently possible.

\section{The Fireball model}
\label{sec:fireball}

\subsection{Relativistic expansion}
\label{sec:Rel-expansion}

General phenomenological considerations, based on $\gamma$-ray
observations, indicate that,
regardless of the nature of the underlying sources, 
GRBs are produced by the dissipation of the kinetic energy of a 
relativistic expanding fireball. 
The rapid rise time and short duration, $\sim1$~ms, 
observed in some bursts \cite{Bhat92,Fishman94} imply
that the sources are compact, with a linear scale comparable
to a light-ms, $r_0\sim10^7$~cm. 
The high $\gamma$-ray luminosity implied by cosmological distances, 
$L_\gamma\sim10^{52}{\rm erg\ s}^{-1}$,
then results in a very high optical depth to pair creation. 
The energy of observed $\gamma$-ray photons is above
the threshold for pair production. The number
density of photons at the source $n_\gamma$ is approximately given by
$L_\gamma=4\pi r_0^2 cn_\gamma\epsilon$, 
where $\epsilon\simeq1$MeV is the characteristic
photon energy. Using $r_0\sim10^7$cm, the optical depth for pair production
at the source is
\begin{equation}
\tau_{\gamma\gamma}\sim r_0 n_\gamma\sigma_T\sim{\sigma_TL_\gamma
\over4\pi r_0 c\epsilon}\sim10^{15},
\label{eq:tau-pair}
\end{equation}
where $\sigma_T$ is the Thomson cross section.

The high optical depth implies that a thermal plasma 
of photons, electrons and positrons is created, a ``fireball,''
which then expands and accelerates to 
relativistic velocities \cite{Bohdan86,Goodman86}. 
The optical depth is reduced by relativistic expansion of the source:
If the source expands with a Lorentz factor $\Gamma$,
the energy of photons in the source frame is smaller by a factor $\Gamma$
compared to that in the observer frame,
and most photons may therefore be below the pair production threshold.

A lower limit for $\Gamma$ may be obtained in the following way 
\cite{Krolik,Baring}. The GRB photon spectrum is well fitted in the
Burst and Transient Source Experiment (BATSE) detectors 
range, 20~keV to 2~MeV \cite{Fishman}, by a combination
of two power-laws, $dn_\gamma/d\epsilon_\gamma\propto\epsilon_\gamma^{-\beta}$ 
with different values of $\beta$ at low and high energy \cite{Band}.
Here, $dn_\gamma/d\epsilon_\gamma$ is the number of photons per unit 
photon energy. The
break energy (where $\beta$ changes) in the observer frame is typically 
$\epsilon_{\gamma b}\sim1{\rm MeV}$, 
with $\beta\simeq1$ at energies below the break and $\beta\simeq2$ 
above the break. In several cases, the spectrum was observed to extend
to energies $>100$~MeV \cite{GRB100MeV,Fishman}.
Consider then a high energy test photon, with observed energy
$\epsilon_t$, trying to escape the relativistically expanding source.
Assuming that in the source rest frame the photon distribution is isotropic,
and that the spectrum of high energy photons follows 
$dn_\gamma/d\epsilon_\gamma\propto\epsilon_\gamma^{-2}$, 
the mean free path for pair
production (in the source rest frame) for a photon of energy 
$\epsilon'_{t}=\epsilon_t/\gamma$ (in the source rest frame) is
\begin{equation}
l_{\gamma\gamma}^{-1}(\epsilon'_{t})
={1\over2}{3\over16}\sigma_T\int d\cos\theta(1-\cos\theta)
\int_{\epsilon_{\rm th}(\epsilon'_{t},\theta)}^\infty d\epsilon
{U_\gamma\over2\epsilon^2}
={1\over16}\sigma_T{U_\gamma\epsilon'_{\rm t}\over(m_ec^2)^2} \,.
\label{eq:lgg}
\end{equation}
Here, $\epsilon_{\rm th}(\epsilon'_{t},\theta)$ is the minimum energy of
photons that may produce pairs interacting with the test photon, 
given by $\epsilon_{\rm th}\epsilon'_t(1-\cos\theta)\ge2(m_ec^2)^2$ 
($\theta$ is the angle between the photons' momentum vectors).
$U_\gamma$ is the photon energy density (in the range
corresponding to the observed BATSE range) in the source rest-frame,
given by $L_\gamma=4\pi r^2\gamma^2cU_\gamma$.
Note , that we have used a constant cross section, $3\sigma_T/16$, 
above the threshold $\epsilon_
{\rm th}$. The cross section drops as $\log(\epsilon)/\epsilon$ for
$\epsilon\gg\epsilon_{\rm th}$; however, since the number density of
photons drops rapidly with energy, this does not introduce a large correction
to $l_{\gamma\gamma}$.

The source size constraint implied by the variability time is modified
for a relativistically expanding source. Since in the observer frame almost 
all photons propagate at a direction making an angle $<1/\Gamma$ with respect 
to the expansion direction, radiation seen by a distant observer
originates from a conical section of the source around the source-observer 
line of sight, with opening angle $\sim1/\Gamma$. Photons which are emitted
from the edge of the cone are delayed, compared to those emitted on the
line of sight, by $r/2\Gamma^2c$. Thus, the constraint on source size
implied by variability on time scale $\Delta t$ is 
\begin{equation}
r\sim2\Gamma^2c\Delta t. 
\label{eq:r_s}
\end{equation}
The time $r/c$ required for significant source expansion
corresponds to comoving time (measured in the source frame)
$t_{\rm co.}\approx r/\Gamma c$. The two-photon collision rate at the source
frame is $t_{\gamma\gamma}^{-1}=c/l_{\gamma\gamma}$. Thus, the source
optical depth to pair production is 
$\tau_{\gamma\gamma}=t_{\rm co.}/t_{\gamma\gamma}\approx
r/\Gamma l_{\gamma\gamma}$. Using Eqs. (\ref{eq:lgg}) and (\ref{eq:r_s}) 
we have
\begin{equation}
\tau_{\gamma\gamma}={1\over128\pi} {\sigma_TL_\gamma\epsilon_t
\over c^2(m_ec^2)^2\Gamma^6\Delta t}. 
\label{eq:taugg}
\end{equation}
Requiring $\tau_{\gamma\gamma}<1$ at $\epsilon_t$ 
we obtain a lower limit for $\Gamma$,
\begin{equation}
\Gamma\ge250\left[
L_{\gamma,52} \left({\epsilon_t
\over100{\rm MeV}}\right)\Delta t_{-2}^{-1}\right]^{1/6},
\label{eq:Gmingg}
\end{equation}
where $L_{\gamma}=10^{52}L_{\gamma,52}{\rm erg/s}$ and 
$\Delta t=10^{-2}\Delta t_{-2}$~s.

\subsection{Fireball evolution}
\label{sec:fireball-evolution}

As the fireball expands it cools, the photon
temperature $T_\gamma$ 
in the fireball frame decreases, and most pairs annihilate.
Once the pair density is sufficiently low, photons may escape.
However, if the observed radiation is due
to photons escaping the fireball as it becomes optically thin, two 
problems arise. First, the photon spectrum is quasi-thermal,
in contrast with observations. Second, 
the source size, $r_0\sim10^7$~cm, and the total energy emitted in 
$\gamma$-rays,
$\sim10^{53}$~erg, suggest that the underlying energy source is related
to the gravitational collapse of $\sim 1 M_\odot$ object.
Thus, the plasma is expected to be ``loaded''
with baryons which may be injected with the radiation or present in the 
atmosphere surrounding the source. A small baryonic load, $\geq10^{-8}
{M_\odot}$, increases the optical depth due to Thomson scattering
on electrons associated with the ``loading'' protons, so that 
most of the radiation energy is converted to
kinetic energy of the relativistically expanding baryons before the plasma
becomes optically thin \cite{kinetic1,kinetic2}. 
To overcome both problems it was proposed \cite{RnM92} that the
observed burst is produced once the kinetic energy of the ultra-relativistic 
ejecta is re-randomized by some dissipation process at large radius, beyond
the Thomson photosphere, and then radiated as $\gamma$-rays. Collision 
of the relativistic baryons
with the inter-stellar medium \cite{RnM92}, and 
internal collisions within the ejecta itself 
\cite{internal1,internal2,internal3}, 
were proposed as possible dissipation processes.
Most GRBs show variability on time scales much shorter than (typically
one hundredth of) the total
GRB duration. Such variability is hard to explain in models where the
energy dissipation is due to external shocks \cite{Woods95,SnP_var}.
Thus, it is believed that internal collisions are responsible for the
emission of $\gamma$-rays. 

Let us first consider the case where the energy release from the source
is ``instantaneous,'' i.e. on a time scale $r_0/c$. We assume that most of 
the energy is released in the form of photons, i.e. that the fraction of 
energy carried by baryon rest mass $M$ satisfies
$\eta^{-1}\equiv Mc^2/E\ll1$. 
The initial thickness of the fireball shell
is $r_0$. Since the plasma accelerates to relativistic
velocity, all fluid elements move with velocity close to $c$, and
the shell thickness remains constant at $r_0$ (this breaks at very late
time, as discussed below).
We are interested in the stage where the optical depth (due to pairs and/or
electrons associated with baryons) is high, but only a small fraction of
the energy is carried by pairs. 

The entropy of a fluid component with zero chemical potential is 
$S=V(e+p)/T$, where $e$, $p$ and $V$ are the (rest frame) energy density, 
pressure and volume. For the photons $p=e/3\propto T_\gamma^4$.
Since initially both the rest mass and thermal energy of
baryons is negligible, the entropy is provided by the photons.
Conservation of entropy implies
\begin{equation}
r^2 \gamma(r) r_0 T_\gamma^3(r)={\rm Constant} ,
\label{eq:entropy}
\end{equation}
and conservation of energy implies
\begin{equation}
r^2 \gamma(r) r_0 \gamma(r) T_\gamma^4(r)={\rm Constant}\,.
\label{eq:energy}
\end{equation}
Here $\gamma(r)$ is the shell Lorentz factor. 
Combining (\ref{eq:entropy}) and (\ref{eq:energy}) we find
\begin{equation}
\gamma(r)\propto r,\quad T_\gamma(r)\propto r^{-1},\quad n\propto r^{-3},
\label{eq:scaling1}
\end{equation}
where $n$
is the rest frame (comoving) baryon number density.

As the shell accelerates the baryon kinetic energy, $\gamma Mc^2$,
increases. It becomes comparable to the total fireball energy when 
$\gamma\sim\eta$, at radius $r_f=\eta r_0$.
At this radius most of the energy of the fireball is carried by the baryon
kinetic energy, and the shell does not accelerate further. 
Eq. (\ref{eq:energy}) describing
energy conservation is replaced with 
$\gamma={\rm Constant}$. Eq. (\ref{eq:entropy}), however, still holds. 
Eq. (\ref{eq:entropy}) 
may be written as $T_\gamma^4/nT_\gamma={\rm Constant}$ (constant entropy per 
baryon). This implies that the
ratio of radiation energy density to thermal energy density associated
with the baryons is $r$ independent.
Thus, the thermal energy associated with the baryons may be neglected at all 
times, and Eq. (\ref{eq:entropy}) 
holds also for the stage where most of the fireball energy 
is carried by the baryon kinetic energy. Thus, for $r>r_f$ we have
\begin{equation}
\gamma(r)=\Gamma\approx\eta,\quad T\propto r^{-2/3},\quad n\propto r^{-2}.
\label{eq:scaling2}
\end{equation}

Let us consider now the case of extended emission from the source, on time
scale $\gg r_0/c$. In this case, the source continuously emits energy
at a rate $L$, and the energy emission is accompanied by mass loss
rate $\dot M=L/\eta c^2$.
For $r<r_f$ the fluid energy density is relativistic,
$aT_\gamma^4/nm_pc^2=\eta r_0/r$, and the speed of sound is $\sim c$. The time
it takes the shell at radius $r$ to expand significantly 
is $r/c$ in the observer frame, 
corresponding to $t_{\rm co.}\sim r/\gamma c$ 
in the shell frame. During this time
sound waves can travel a distance 
$cr/\gamma c$ in the shell frame, corresponding to $r/\gamma^2=r/(r/r_0)^2=
(r_0/r)r_0$ in the observer frame. This implies that at the early stages of
evolution, $r\sim r_0$, sound waves had enough time to smooth out spatial
fluctuations in the fireball over a scale $r_0$, but that regions separated by
$\Delta r>r_0$ can not interact with each other. Thus, if the emission extends
over a time $T_{\rm GRB}\gg r_0/c$, 
a fireball of thickness $cT_{\rm GRB}\gg r_0$ would be formed,
which would expand as a collection of independent, roughly uniform, sub-shells
of thickness $r_0$. Each sub-shell would reach a final Lorentz factor 
$\Gamma_f$, which may vary between sub-shells. This
implies that different sub-shells may have velocities differing by 
$\Delta v\sim c/2\eta^2$, where $\eta$ is some typical value representative
of the entire fireball. Different shells emitted at times differing
by $\Delta t$, $r_0/c<\Delta t<T_{\rm GRB}$, may therefore collide with
each other after a time $t_c\sim c\Delta t/\Delta v$, i.e. at a radius
\begin{equation}
r_i\approx2\Gamma^2c\Delta t=6\times10^{13}\Gamma^2_{2.5}\Delta t_{-2}
{\rm\ cm},
\label{eq:r_i}
\end{equation}
where $\Gamma=10^{2.5}\Gamma_{2.5}$.
The minimum internal shock radius, $r\sim \Gamma^2 r_0$, 
is also the radius at which an individual sub-shell may experience
significant change in its width $r_0$, due to Lorentz factor variation
across the shell.

\subsection{The allowed range of Lorentz factors and baryon loading}
\label{sec:eta}

The acceleration, $\gamma\propto r$, of fireball plasma
is driven by radiation pressure. Fireball
protons are accelerated through their coupling to the electrons, which
are coupled to fireball photons. 
We have assumed in the analysis presented above, that photons and electrons
are coupled throughout the acceleration phase. However, if the baryon loading
is too low, radiation may decouple from fireball electrons 
already at $r<r_f$. The fireball
Thomson optical depth is given by the product of comoving expansion time,
$r/\gamma(r) c$, and the photon Thomson scattering rate, $n_ec\sigma_T$.
The electron and proton comoving number densities are
equal, $n_e=n_p$, and are determined by equating the $r$ independent
mass flux carried by the wind, $4\pi r^2 c \gamma(r) n_p m_p$, 
to the mass loss rate from the underlying 
source, which is related to the rate $L$ at which energy 
is emitted through $\dot M=L/(\eta c^2)$. Thus,
during the acceleration phase, $\gamma(r)=r/r_0$, the 
Thomson optical depth $\tau_T\propto r^{-3}$. $\tau_T$ drops 
below unity at a radius $r<r_f=\eta r_0$ if $\eta>\eta_*$, where
\begin{equation}
\eta_*=\left({\sigma_T L \over 4\pi r_0 m_p c^3}\right)^{1/4}=
       1.0\times10^3 L_{52}^{1/4}r_{0,7}^{-1/4}.
\label{eq:eta-star}
\end{equation}
Here $r_0=10^7r_{0,7}$~cm.

If $\eta>\eta_*$ radiation decouples from the fireball plasma at
$\gamma=r/r_0=\eta_*^{4/3}\eta^{-1/3}$. If $\eta\gg\eta_*$, then most of
the radiation energy is not converted to kinetic energy prior to radiation
decoupling, and most of the fireball energy escapes in the form of thermal
radiation. Thus, the baryon load of fireball shells, and the corresponding 
final Lorentz factors, must be within the range 
$10^2\le\Gamma\approx\eta\le\eta_*\approx10^3$ 
in order to allow the production of the
observed non-thermal $\gamma$-ray spectrum.

\subsection{Fireball interaction with surrounding medium}
\label{sec:fireball-interaction}

As the fireball expands, it drives a relativistic shock (blast-wave)
into the surrounding
gas, e.g. into the inter-stellar medium (ISM) gas if the explosion 
occurs within a galaxy. In what follows, we refer to the surrounding gas
as ``ISM gas,'' although the gas need not necessarily be inter-stellar.
At early time, the fireball is little affected by the interaction with the 
ISM. At late time, most of the fireball energy is transferred to the ISM, and
the flow approaches the self-similar blast-wave solution of Blandford \&
McKee \cite{BnM76}. At this stage a single shock propagates into the ISM,
behind which the gas expands with Lorentz factor
\begin{equation}
\Gamma_{BM}(r)=\left(\frac{17E}{16\pi n m_p c^2}\right)^{1/2}r^{-3/2}=
150\left({E_{53}\over n_0}\right)^{1/2}r_{17}^{-3/2},
\label{eq:Gamma_BM}
\end{equation}
where $E=10^{53}E_{53}$~erg is the fireball energy,
$n=1n_0{\rm\ cm}^{-3}$ is the ISM number density, and $r=10^{17}r_{17}$~cm
is the shell radius. 
The characteristic time
at which radiation emitted by shocked plasma at radius $r$
is observed by a distant observer is $t\approx r/4\Gamma_{BM}^2 c$
\cite{WAG-ring}.

The transition to self-similar expansion occurs on a time scale $T$ (measured
in the observer frame)
comparable to the longer of the two time scales set by the initial
conditions: the (observer) GRB duration $T_{\rm GRB}$ and the
(observer) time $T_{\Gamma}$ at which the self-similar Lorentz factor
equals the original ejecta Lorentz factor $\Gamma$,
$\Gamma_{BM}(t=T_\Gamma)=\Gamma$. Since $t=r/4\Gamma^2_{BM}c$,
\begin{equation}
T=\max\left[T_{\rm GRB}, 5\left({E_{53}\over
n_0}\right)^{1/3}\Gamma_{2.5}^{-8/3}{\rm\, s}\right].
\label{eq:t_tr}
\end{equation}
During the transition,  plasma shocked by the reverse shocks
expands with Lorentz factor close to that given by the self-similar solution,
\begin{equation}
\Gamma_{\rm tr.}\simeq\Gamma_{BM}(t=T)
\simeq 245\left({E_{53}\over n_0}\right)^{1/8}T_1^{-3/8},
\label{eq:Gamma_tr}
\end{equation}
where $T=10T_1$~s. The unshocked fireball
ejecta propagate at the original expansion Lorentz factor, 
$\Gamma$,
and the Lorentz factor of plasma shocked by the reverse shock in
the rest frame of the unshocked ejecta is $\simeq\Gamma/\Gamma_{\rm tr.}$.
If $T\simeq T_{\rm GRB}\gg T_{\Gamma}$ then 
$\Gamma/\Gamma_{\rm tr.}\gg1$, the reverse shock is relativistic,
and the Lorentz factor associated with the random motion of protons 
in the reverse shock is $\gamma_p^R\simeq\Gamma/\Gamma_{\rm tr.}$. 

If, on the other hand, $T\simeq T_{\Gamma}\gg T_{\rm GRB}$ then 
$\Gamma/\Gamma_{\rm tr.}\sim1$, and the reverse shock is not relativistic. 
Nevertheless, the following argument suggests that the reverse shock
speed is not far below $c$, and that the protons are therefore 
heated to 
relativistic energy, $\gamma_p^R-1\simeq1$. The comoving time, measured
in the fireball ejecta frame prior to deceleration, 
is $t_{\rm co.}\simeq r/\Gamma c$. The expansion Lorentz
factor is expected to vary across the ejecta, $\Delta \Gamma/\Gamma\sim1$,
due to variability of the underlying GRB source over the duration of 
its energy release.
Such variation would lead
to expansion of the ejecta, in the comoving frame, at relativistic speed. 
Thus, at the deceleration radius, $t_{\rm co.}\simeq\Gamma T$,
the ejecta width exceeds $\simeq ct_{\rm co.}\simeq\Gamma c T$. Since
the reverse shock should cross the ejecta over a deceleration time 
scale, $\simeq\Gamma T$, the reverse shock speed must be close to 
$c$. We therefore conclude that 
the Lorentz factor associated with the random motion of protons 
in the reverse shock is approximately given by 
$\gamma_p^R-1\simeq \Gamma/\Gamma_{\rm tr.}$ for both 
$\Gamma/\Gamma_{\rm tr.}\sim1$
and $\Gamma/\Gamma_{\rm tr.}\gg1$. 

Since $T_{\rm GRB}\sim10$~s is typically comparable to $T_\Gamma$, 
the reverse shocks are typically expected to be mildly relativistic.

\subsection{Fireball geometry}
\label{sec:fireball-geometry}

We have assumed in the discussion so far that the fireball is spherically 
symmetric. However, a jet-like fireball behaves as if it were
a conical section of a spherical fireball as long as the jet opening
angle is larger than $\Gamma^{-1}$. This is due to the fact that
the linear size of causally connected regions, 
$ct_{\rm co.}\sim r/\Gamma$ in the fireball frame, corresponds to
an angular size $ct_{\rm co.}/r\sim\Gamma^{-1}$. Moreover, due
to the relativistic beaming of radiation, a distant observer can not
distinguish between a spherical fireball and a jet-like fireball, as
long as the jet opening angle $\theta>\Gamma^{-1}$. Thus, as long as 
we are discussing processes that occur when
the wind is ultra-relativistic, $\Gamma\sim300$ (prior to 
significant fireball deceleration by the surrounding medium), our
results apply for both a spherical and a jet-like fireball.
In the latter case, $L$ ($E$) in our
equations should be understood as the luminosity (energy) the fireball
would have carried had it been spherically symmetric.

\subsection{$\gamma$-ray emission}
\label{sec:fireball-gammas}

If the Lorentz factor variability within the wind is significant,
internal shocks would reconvert a substantial 
part of the kinetic energy to internal energy. The internal 
energy may then be radiated as 
$\gamma$-rays by synchrotron and inverse-Compton emission of
shock-accelerated electrons. The
internal shocks are expected to be ``mildly'' relativistic in the fireball 
rest frame, i.e. characterized by Lorentz factor 
$\Gamma_i-1\sim$ a few. This is due to the fact that the allowed range
of shell Lorentz factors is $\sim10^2$ to $\sim10^3$ (see \S\ref{sec:eta}),
implying that the Lorentz factors associated with the relative velocities
are not very large. Since internal shocks are mildly
relativistic, we expect results related to particle
acceleration in sub-relativistic shocks (see \cite{Blandford87} for review)
to be valid for acceleration
in internal shocks. In particular, electrons are 
expected to be accelerated to a power law energy distribution,
$dn_e/d\gamma_e\propto\gamma_e^{-p}$ for $\gamma_e>\gamma_m$, with $p\simeq2$
\cite{AXL77,Bell78,BnO78}.

The minimum Lorentz factor $\gamma_m$
is determined by the following consideration. 
Protons are heated in internal shocks
to random velocities (in the wind frame) 
$\gamma_p^R-1\approx\Gamma_i-1\approx1$. 
If electrons carry a fraction $\xi_e$ of the shock internal energy, then
$\gamma_m\approx\xi_e(m_p/m_e)$. The characteristic
frequency of synchrotron emission is determined by $\gamma_m$ and 
by the strength of the magnetic field. Assuming 
that a fraction $\xi_B$ of the internal energy 
is carried by the magnetic field,  
$4\pi r_i^2c\Gamma^2B^2/8\pi=\xi_B L_{\rm int.}$, 
the characteristic observed energy of synchrotron photons, 
$\epsilon_{\gamma b}=\Gamma\hbar\gamma_m^2 eB/m_ec$, is
\begin{equation}
\epsilon_{\gamma b}\approx1\xi_B^{1/2}\xi_e^{3/2}{L_{\gamma,52}^{1/2}
\over\Gamma_{2.5}^2\Delta t_{-2}}{\rm MeV}.
\label{eq:E_gamma}
\end{equation}
In deriving Eq. (\ref{eq:E_gamma}) we have assumed that the
wind luminosity carried by internal plasma energy, $L_{\rm int.}$, is
related to the observed $\gamma$-ray luminosity through 
$L_{\rm int.}=L_\gamma/\xi_e$. This assumption is justified since
the electron synchrotron cooling time is short
compared to the wind expansion time (unless the equipartition fraction 
$\xi_B$ is many orders of magnitude smaller than unity), and 
hence electrons lose all their energy radiatively. 
Fast electron cooling also results in a synchrotron
spectrum $dn_\gamma/d\epsilon_\gamma\propto\epsilon_\gamma^{-1-p/2}=
\epsilon_\gamma^{-2}$ at $\epsilon_\gamma>\epsilon_{\gamma b}$,
consistent with observed GRB spectra \cite{Band}.

At present, there is no theory that allows the determination of 
the values of the equipartition fractions $\xi_e$ and $\xi_B$. 
Eq. (\ref{eq:E_gamma}) implies that 
fractions not far below unity are required to account for the observed
$\gamma$-ray emission. We note, that build up of magnetic field
to near equipartition by electro-magnetic instabilities is expected to
be a generic characteristic of collisionless shocks 
(see discussion in ref. \cite{Blandford87} and references therein), and
is inferred to occur in other systems, e.g. in supernova remnant shocks
(e.g. \cite{Helfand87,Cargill88}).

The $\gamma$-ray break energy $\epsilon_{\gamma b}$
of most GRBs 
observed by BATSE detectors is in the range of
100~keV to 300~keV \cite{Brainerd99}. It may appear from
Eq. (\ref{eq:E_gamma}) that the
clustering of break energies in this narrow energy range requires
fine tuning of fireball model parameters, which should naturally 
produce a much wider range of break energies. 
This is, however, not the case \cite{GSW00}. Consider the dependence of 
$\epsilon_{\gamma b}$ on $\Gamma$. The strong $\Gamma$ dependence of
the pair-production optical depth, Eq. (\ref{eq:taugg}), implies
that if the value of $\Gamma$ is smaller than the minimum value allowed 
by Eq. (\ref{eq:Gmingg}), for which 
$\tau_{\gamma\gamma}(\epsilon_\gamma=100{\rm MeV})\approx1$,
most of the high energy photons in the power-law distribution
produced by synchrotron emission, 
$dn_\gamma/d\epsilon_\gamma\propto\epsilon_\gamma^{-2}$, would be converted
to pairs. This would lead to high optical depth due to Thomson scattering 
on $e^\pm$, and hence to strong suppression of the emitted flux
\cite{GSW00}. For
fireball parameters such that 
$\tau_{\gamma\gamma}(\epsilon_\gamma=100{\rm MeV})\approx1$, the break energy
implied by Eqs. (\ref{eq:E_gamma}) and  (\ref{eq:Gmingg}) is
\begin{equation}
\epsilon_{\gamma b}\approx1\xi_B^{1/2}\xi_e^{3/2}{L_{\gamma,52}^{1/6}
\over\Delta t_{-2}^{2/3}}{\rm MeV}.
\label{eq:E_gamma-max}
\end{equation}
As explained in \S\ref{sec:eta}, shell Lorentz factors can not exceed
$\eta_*\simeq10^3$, for which break energies in the X-ray range,
$\epsilon_{\gamma b}\sim10$~keV, may be obtained. We note,
however, that the radiative flux would be strongly 
suppressed in this case too \cite{GSW00}. 
If the typical $\Gamma$ of
radiation emitting shells is close to $\eta_*$, then the 
range of Lorentz factors of wind shells is narrow,
which implies that only a small fraction of wind kinetic
energy would be converted to internal energy which can be radiated from the
fireball.

Thus, the clustering of break energies at $\sim1$~MeV is naturally accounted 
for, provided that the variability time scale satisfies
$\Delta t\le10^{-2}$~s, which implies an upper limit on the source size,
since $\Delta t\ge r_0/c$. We note, that 
a large fraction of bursts detected by BATSE show variability
on the shortest resolved time scale, $\sim10$~ms \cite{Woods95}. 
In addition, a natural
consequence of the model is the existence of low luminosity
bursts with low, 1~keV to 10~keV, break energies \cite{GSW00}. Such 
``X-ray bursts'' may have recently been identified \cite{FXTs}.

For internal collisions, the observed
$\gamma$-ray variability time, $\sim r_i/\Gamma^2 c\approx\Delta t$,
reflects the variability time of the underlying source, and the GRB
duration reflects the duration over which energy is emitted from the
source. Since the wind Lorentz factor is expected to fluctuate on
time scales ranging from the shortest variability time $r_0/c$ to the
wind duration $T_{\rm GRB}$, internal collisions will take place over a range
of radii, $r\sim\Gamma^2 r_0$ to $r\sim\Gamma^2cT_{\rm GRB}$.

\subsection{Afterglow emission}
\label{sec:fireball-afterglow}

Let us consider the radiation emitted from the reverse shocks,
during the transition to self-similar expansion.
The characteristic electron
Lorentz factor (in the plasma rest frame)
is $\gamma_m\simeq\xi_e(\Gamma/\Gamma_{\rm tr.})m_p/m_e$, where
the internal energy per proton in the shocked ejecta
is $\simeq (\Gamma/\Gamma_{\rm tr.})m_p c^2$. The energy density
$U$ is  $E\approx 4\pi r^2cT \Gamma_{\rm tr.}^2 U$,
and the number of radiating electrons is $N_e\approx E/\Gamma m_p c^2$.
Using Eq. (\ref{eq:Gamma_tr}) and $r=4\Gamma_{\rm tr.}^2cT$, the 
characteristic (or peak) energy of
synchrotron photons (in the observer frame) is \cite{Draine00}
\begin{equation}
\epsilon_{\gamma m}
\approx\hbar\Gamma_{\rm tr.}\gamma_m^2{eB\over m_e c}=
2\xi_{e,-1}^2\xi_{B,-1}^{1/2}n_0^{1/2}
\Gamma_{2.5}^{2}{\rm\, eV},
\label{eq:e_m}
\end{equation}
and the specific luminosity,
$L_\epsilon=dL/d\epsilon_\gamma$, at
$\epsilon_{\gamma m}$ is
\begin{eqnarray}
L_m&&\approx
(2\pi\hbar)^{-1}\Gamma_{\rm tr.}{e^3B\over m_e c^2}N_e\approx
10^{61}\xi_{B,-1}^{1/2}E_{53}^{5/4}T_1^{-3/4}
\Gamma_{2.5}^{-1}n_0^{1/4}{\rm\, s}^{-1},
\label{eq:L_m}
\end{eqnarray}
where $\xi_e=0.1\xi_{e,-1}$, and $\xi_B=0.1\xi_{B,-1}$.

Here too, we expect a power law energy distribution,
$dN_e/d\gamma_e\propto\gamma_e^{-p}$ for $\gamma_e>\gamma_m$, with $p\simeq2$.
Since the radiative cooling time of electrons in the reverse shock
is long compared to the ejecta expansion time, 
the specific luminosity extends in this case to energy
$\epsilon_\gamma>\epsilon_{\gamma m}$ as
$L_\epsilon=L_m(\epsilon_\gamma/\epsilon_{\gamma m})^{-1/2}$,
up to photon energy
$\epsilon_{\gamma c}$. Here $\epsilon_{\gamma c}$ is
the characteristic synchrotron frequency of
electrons for which the synchrotron cooling time,
$6\pi m_e c/\sigma_T\gamma_e B^2$, is comparable to the ejecta (rest frame)
expansion time, $\sim \Gamma_{\rm tr.} T$. At energy
$\epsilon_\gamma>\epsilon_{\gamma c}$,
\begin{equation}
\epsilon_{\gamma c}\approx
0.1\xi_{B,-1}^{-3/2}n_0^{-1}E_{53}^{-1/2}T_1^{-1/2}
{\rm\, keV},
\label{eq:e_c}
\end{equation}
the spectrum steepens to $L_\epsilon\propto\epsilon_\gamma^{-1}$.

The shock driven into the ISM continuously heats new gas, and
produces relativistic 
electrons that may produce the delayed afterglow radiation
observed on time scales $t\gg T$, typically of order
days to months. As the shock-wave decelerates, the emission shifts to
lower frequency with time. Since we are interested in proton acceleration 
to high energy and in the production of high energy neutrinos, which  take
place primarily in the internal and reverse shocks 
(see \S\ref{sec:UHECR}, \S\ref{sec:nus}), we do
not discuss in detail the theory of late-time afterglow emission.

\section{Some implications of afterglow observations}
\label{sec:afterglow}

Afterglow observations lead to the confirmation, 
as mentioned in the Introduction, of 
the cosmological origin of GRBs \cite{AG_ex_review}, 
and confirmed \cite{Waxman97a,Wijers97} standard model
predictions \cite{Rhoads93,Katz94,MnR97,Vietri97a}
of afterglow that results from 
synchrotron emission of electrons accelerated to
high energy in the highly relativistic shock driven by the fireball into
its surrounding gas. 
Since we are interested mainly in the earlier, internal
collision phase of fireball evolution, we do not discuss afterglow 
observations in detail. We note, however, several implications of
afterglow observations which are of importance for the discussion of UHECR
production. 

The following 
point should be clarified in the context of afterglow observations.
The distribution of GRB durations is bimodal, with broad peaks at
$T_{\rm GRB}=0.2$~s and $T_{\rm GRB}=20$~s \cite{Fishman}. The
majority of bursts belong to the long duration, $T_{\rm GRB}\sim20$~s,
class. The detection
of afterglow emission was made possible thanks to the accurate GRB
positions provided on hour time scale by the 
{\it BeppoSAX} satellite \cite{Costa97}. Since the detectors on board
this satellite trigger only on long bursts, afterglow observations are
not available for the sub-population of short, $T_{\rm GRB}\sim0.2$~s,
bursts. Thus, while the discussion of the fireball model presented in
\S\ref{sec:fireball}, based on $\gamma$-ray observations and on simple
phenomenological arguments, applies to both long and short 
duration bursts, 
the discussion below of afterglow observations applies to long duration
bursts only. It should therefore be kept in mind that short duration bursts
may constitute a different class of GRBs, which, for example,
may be produced by a 
different class of progenitors and may have a different redshift distribution
than the long duration bursts.

Prior to the detection of afterglows, it was commonly 
assumed that the farthest observed GRBs
lie at redshift $z\sim1$. Following the detection of afterglows
and the determination of GRB redshifts, it is now clear that most
GRB sources lie within the redshift range $z\sim0.5$ to $z\sim2$, with 
some bursts observed at $z>3$. For the average GRB $\gamma$-ray fluence,
$1.2\times10^{-5}{\rm erg/cm}^2$ in the 20~keV to 2~MeV band, 
this implies characteristic isotropic $\gamma$-ray energy and luminosity
$E_\gamma\sim10^{53}$~erg and $L_\gamma\sim10^{52}{\rm\ erg/s}$
(in the 20~keV to 2~MeV band), about an
order of magnitude higher than the values assumed prior to 
afterglow detection (Here, and throughout the paper we assume a flat universe
with $\Omega=0.3$, $\Lambda=0.7$, and $H_0=65{\rm km/s\,Mpc}$). 
These estimates are consistent with more detailed analyses of the 
GRB luminosity function and redshift distribution. Mao \& Mo, e.g.,
find, for the cosmological parameters we use, 
a median GRB energy of $\approx0.6\times10^{53}{\rm erg}$ in the 50keV
to 300keV band \cite{MnM98}, corresponding to a median GRB energy 
of $\approx2\times10^{53}{\rm erg}$ in
the 20~keV to 2~MeV band.

The determination of GRB redshifts also lead to a modification of
GRB rate estimates. Since most observed GRB sources lie within the 
redshift range $z\sim0.5$ 
to $z\sim2$, observations
essentially determine the GRB rate per unit volume
at $z\sim1$. The observed rate of $10^3/{\rm yr}$ implies 
$R_{\rm GRB}(z=1)\approx3/{\rm Gpc}^3{\rm yr}$. 
The present, $z=0$, rate is less well constrained, since available data
are consistent with both no evolution of GRB rate with redshift, and 
with strong evolution (following, e.g.,
star formation rate), in which $R_{\rm GRB}(z=1)/R_{\rm GRB}(z=0)\sim10$
\cite{GRB_z1,GRB_z2}.
Detailed analyses, assuming $R_{\rm GRB}$ is proportional to star formation 
rate, lead to $R_{\rm GRB}(z=0)\sim0.5/{\rm Gpc}^3{\rm yr}$ 
\cite{AG_ex_review}. The implied local ($z=0$) $\gamma$-ray
energy generation rate by GRBs in the 20~keV to 2~MeV band is therefore
\begin{equation}
\dot\varepsilon_\gamma(z=0)=10^{44}\zeta\,{\rm erg/Mpc}^3{\rm yr},
\label{eq:gamma_rate}
\end{equation}
with $\zeta$ in the range of $\sim10^{-0.5}$ to $\sim10^{0.5}$.
Note, that $\dot\varepsilon_\gamma$ is independent of the fireball geometry.
If fireballs are conical jets of solid angle $\Delta\Omega$, then the
total energy released by each burst is smaller by a factor
$\Delta\Omega/4\pi$ than the isotropic energy, and the GRB rate is larger
by the same factor.

Due to present technical limitations of the experiments, 
afterglow radiation is observed in most cases only on time scale $>>10$~s.
At this stage, radiation is produced by the
external shock driven into the surrounding gas, and afterglow observations
therefore do not provide direct constraints on plasma parameters
at the internal and reverse shocks, where
protons are accelerated to ultra-high energy.
In one case, however, 
that of GRB~990123, optical emission has been detected on $\sim10$~s
time scale \cite{Akerlof99}. The most natural explanation of the 
observed optical radiation is synchrotron emission from electrons accelerated
to high energy in the reverse shocks driven into fireball ejecta at the
onset of interaction with the surrounding medium \cite{MR_0123,SP_0123},
as explained in \S\ref{sec:fireball-afterglow}.
This observation provides therefore direct constraints on the fireball
ejecta plasma. First, it provides
strong support for one of the underlying assumptions of the 
dissipative fireball scenario described in 
\S\ref{sec:fireball-evolution}, that the energy is carried from the 
underlying source in the form of proton kinetic energy. This is due to the
fact that the observed radiation is well accounted for in a model
where a shock propagates
into fireball plasma composed of protons and electrons (rather than,
e.g., pair plasma). Second, 
comparison of the observed flux with model predictions, Eqs. 
(\ref{eq:e_m}) and (\ref{eq:L_m}), implies
$\xi_e\sim\xi_B\sim10^{-1}$. 

Afterglow observations imply that a significant fraction
of the energy initially carried by the fireball is converted into 
$\gamma$-rays, i.e. that the observed $\gamma$-ray energy provides a rough
estimate of the total fireball energy. This has been demonstrated for
one case, that of GRB970508, by a comparison of the total fireball energy
derived from long term radio observations with the energy emitted in 
$\gamma$-rays \cite{WKF98,FWK00}, and for a large number of bursts
by a comparison of observed $\gamma$-ray energy with 
the total fireball energy estimate based on X-ray afterglow data 
\cite{Freedman}. In the context of the fireball
model described in \S\ref{sec:fireball}, 
the inferred high radiative efficiency implies that 
a significant fraction of the wind kinetic energy must be converted to 
internal energy in internal shocks, and that electrons must carry a 
significant fraction of the internal energy, i.e. that $\xi_e$ should
be close to unity. We have already shown, see \S\ref{sec:fireball-gammas},
that $\xi_e$ values not far below unity are required to account for the
observed $\gamma$-ray emission. Conversion in internal shocks
of a large fraction of
fireball kinetic energy to internal energy is possibly provided the
variance in the Lorentz factors of fireball shells is large \cite{GSW00}.

In accordance with the implications of afterglow observations, we assume 
in the discussion below of UHECR and neutrino production in GRB fireballs,
that the fraction of fireball energy converted to internal energy
carried by electrons, and hence to $\gamma$-rays, is large. For discussion of
UHECR production under the assumption that only a negligible fraction,
$\sim m_e/m_p$, of
the fireball energy is converted to radiation see ref. \cite{Totani00}.

\section{UHECRs from GRB fireballs}
\label{sec:UHECR}

\subsection{Fermi acceleration in GRBs}
\label{sec:fermi}

In the fireball model, the observed radiation is produced, both during
the GRB and the afterglow, by synchrotron emission of shock accelerated
electrons. In the region where electrons are accelerated, 
protons are also expected to be
shock accelerated. This is similar to what is thought to occur in supernovae 
remnant shocks, where synchrotron radiation of accelerated electrons is the
likely source of non-thermal X-rays (recent ASCA observations give evidence
for acceleration of electrons in the remnant of SN1006 to $10^{14}{\rm eV}$ 
\cite{SN1006}), and where shock acceleration of protons is believed to
produce cosmic rays with energy extending to $\sim10^{15}{\rm eV}$ (see, e.g.,
\cite{Blandford87} for review). Thus, it is likely that protons, as well
as electrons, are accelerated to high energy within GRB fireballs.
Let us consider the constraints that should be satisfied
by the fireball parameters in order to allow acceleration of protons to $\sim
10^{20}$~eV.

We consider proton Fermi acceleration in fireball internal shocks, which
take place as the fireball expands over a range
of radii, $r\sim\Gamma^2 r_0$ to $r\sim\Gamma^2cT_{\rm GRB}$,
and at the reverse shocks driven into fireball ejecta
due to interaction with surrounding medium at 
$r\sim\Gamma^2cT\sim\Gamma^2cT_{\rm GRB}$ 
(see \S\ref{sec:fireball-evolution},\S\ref{sec:fireball-interaction}).
Both internal and reverse shocks are, in the wind rest-frame,
mildly relativistic, i.e. characterized by Lorentz factors 
$\Gamma_i-1\sim1$. Moreover, since reverse shocks do not
cause strong deceleration of fireball plasma, see 
\S\ref{sec:fireball-interaction}, the expansion Lorentz factor 
$\Gamma_{\rm tr.}$ of fireball plasma 
shocked by reverse shocks is similar to the fireball Lorentz factor $\Gamma$
prior to interaction with the surrounding medium. Thus,    
plasma parameters, e.g. energy and number density, in the reverse shocks 
are similar to those obtained in internal shocks due to variability
on time scale $\Delta t\sim T$. Results obtained below for internal
shocks are therefore valid also for reverse shocks, provided
$\Delta t$ is replaced with $T$.

Since the shocks we are interested in are mildly
relativistic, we expect results related to particle
acceleration in sub-relativistic shocks 
(see \cite{Blandford87} for review) to be valid for our
scenario. The predicted energy distribution of accelerated
protons is therefore $dn_p/d\epsilon_p\propto \epsilon_p^{-2}$
\cite{AXL77,Bell78,BnO78}, similar to the predicted
electron energy spectrum, which is consistent with 
the observed photon spectrum (see \S\ref{sec:fireball-gammas}).

The most restrictive requirement, which rules out the possibility of 
accelerating 
particles to energy $\sim10^{20}$~eV in most astrophysical objects, 
is that
the particle Larmor radius $R_L$ should be smaller than the system size
\cite{Hillas84}. In our scenario we must apply a more stringent requirement,
namely that $R_L$ should be smaller than the largest scale $l$ over which
the magnetic field fluctuates, since otherwise 
Fermi acceleration
may not be efficient. We may estimate $l$ as follows. The comoving time, i.e.
the time measured in the wind rest frame, is $t=r/\Gamma c$. Thus, 
regions separated by a comoving distance larger than $r/\Gamma$ are
causally disconnected, and the wind properties fluctuate over comoving
length scales up to $l\sim r/\Gamma$. We must therefore 
require $R_L<r/\Gamma$.
A somewhat more stringent requirement is related to the wind expansion.
Due to expansion the internal energy is decreasing and therefore
available for proton acceleration (as well as for $\gamma$-ray production) only
over a comoving time $t\sim r/\Gamma c$. The typical Fermi
acceleration time 
is $t_a=f R_L/c\beta^2$ \cite{Hillas84,Blandford87}, 
where $\beta c$ is the Alfv\'en velocity and 
$f\sim 1$ \cite{Kulsrud}. 
In our scenario $\beta\simeq1$ leading to the requirement $fR_L<r/\Gamma$.
This condition sets a lower limit to the required
comoving magnetic field strength. Using the relations 
$R_L=\epsilon'_p/eB=\epsilon_p/\Gamma e B$, where 
$\epsilon'_p=\epsilon_p/\Gamma$ is the proton energy measured in the
fireball frame, and $4\pi r^2c\Gamma^2B^2/8\pi=\xi_B L_\gamma/\xi_e$, 
the constraint $fR_L<r/\Gamma$ may be written as \cite{W95a},
\begin{equation}
\xi_B/\xi_e>0.02 f^2\Gamma_{2.5}^2 \epsilon_{p,20}^2L_{\gamma,52}^{-1},
\label{eq:xiB}
\end{equation}
where $\epsilon_p=10^{20}\epsilon_{p,20}$~eV is the accelerated proton energy. 
Note, that this constraint is independent of $r$, the internal collision
radius.

The accelerated proton energy is also limited by energy loss
due to synchrotron radiation and interaction with fireball photons.
As discussed in \S6, the dominant energy loss process is synchrotron cooling. 
The condition that the synchrotron loss time, $t_{sy}=(6\pi m_p^4 c^3/\sigma_T
m_e^2)\epsilon_p^{-1}B^{-2}$, should be smaller than the acceleration time sets
an upper limit to the magnetic field strength. Since the equipartition field
decreases with radius, 
$B_{e.p.}\propto r^{-2}$, the upper limit on the magnetic
field may be satisfied 
simultaneously with (\ref{eq:xiB}) provided
that the internal collisions occur at large enough radius \cite{W95a},
\begin{equation}
r>10^{12}f^2\Gamma_{2.5}^{-2}\epsilon_{p,20}^3{\rm cm}.
\label{eq:rmin}
\end{equation}
Since collisions occur at radius $r\approx\Gamma^2c\Delta t$, the
condition (\ref{eq:rmin}) is equivalent to a lower limit on $\Gamma$
\begin{equation}
\Gamma>130 f^{1/2} \epsilon_{p,20}^{3/4}\Delta t^{-1/4}_{-2}.
\label{eq:Gmin}
\end{equation}

From Eqs. (\ref{eq:xiB}) and (\ref{eq:Gmin}), we infer that 
a dissipative ultra-relativistic wind,
with luminosity and variability time implied by GRB observations,
satisfies the constraints necessary to allow the acceleration of protons 
to energy $>10^{20}$~eV, provided that the wind bulk Lorentz factor is
large enough, $\Gamma>100$, and that the
magnetic field is close to equipartition with electrons. The former 
constraint, $\Gamma>100$, is remarkably similar to that inferred based on
the $\gamma$-ray spectrum, and $\Gamma\sim300$ is the ``canonical'' 
value assumed in the fireball model. The latter constraint, magnetic field
close to equipartition, must be satisfied to account for
both $\gamma$-ray emission (see \S\ref{sec:fireball-gammas})
and afterglow observations (see \S\ref{sec:afterglow}).

Finally, two points should be clarified. First, it has recently been claimed
that ultra-high energy protons would lose most of their energy adiabatically, 
i.e. due to expansion, before they escape the fireball 
\cite{RnM_adiabatic}. This claim is based on the assumptions that
internal shocks, and therefore proton acceleration, occur at 
$r\sim\Gamma^2r_0$ only, 
and that subsequently the fireball expands
adiabatically. Under these assumptions, protons 
would lose most their energy by the time they escape. 
However, as emphasized both in this section and
in \S\ref{sec:fireball-evolution}, 
internal shocks are expected to occur over a wide range of radii,
and in particular at $r\sim\Gamma^2cT$ during the transition
to self-similar expansion. Thus, proton acceleration to ultra-high energy
is expected to operate over a wide range of radii,
from $r\sim\Gamma^2r_0$ up to $r\sim\Gamma^2cT$, where ultra-high
energy particles escape. 

Second, it has recently been claimed in \cite{Gallant98} that the
conditions at the external shock driven by the fireball into the ambient
gas are not likely to allow proton acceleration to ultra-high energy. 
Regardless of the validity of this claim, it
is irrelevant for the acceleration in internal shocks,
the scenario considered for UHECR production in GRBs in both \cite{W95a} 
and \cite{Vietri95}. Moreover, it is not at all clear that UHECRs can not
be produced at the external shock, since the magnetic field may be 
amplified ahead
of the shock by the streaming of high energy particles.
For discussion of high energy proton production in the external shock and
its possible implications see ref. \cite{Dermer00}.

\subsection{UHECR flux and spectrum}
\label{sec:CRflux}

Fly's Eye \cite{Bird93,Bird94} and AGASA 
\cite{Hayashida94,Yoshida95,Takeda98} results
confirm the flattening of the cosmic-ray spectrum at $\sim10^{19}$~eV,
evidence for which existed in previous experiments with weaker statistics
\cite{Watson91}. Fly's Eye data is well fitted in the energy range 
$10^{17.6}$~eV to $10^{19.6}$~eV by a sum of two power laws: A
steeper component, with differential number spectrum
$J\propto E^{-3.50}$, dominating at lower
energy, and a shallower component, $J\propto E^{-2.61}$, 
dominating at higher energy, $E>10^{19}$~eV.
The flattening of the spectrum, combined with the lack of anisotropy 
and the evidence for a change in composition from heavy nuclei at low
energy to light nuclei (protons) at high energy 
\cite{Watson91,Bird93,Bird94,composition1,composition2},
suggest that a Galactic component of heavy nuclei, $J\propto E^{-3.50}$,
dominates the cosmic-ray flux at low energy, while 
an extra-Galactic component of protons, 
$J\propto E^{-2.61}$, dominates the flux at high energy, $>10^{19}$~eV.

The GRB energy observed in $\gamma$-rays reflects the fireball
energy in accelerated electrons. If 
accelerated electrons and protons carry similar energy, 
as indicated by afterglow observations
\cite{Freedman} (see \S\ref{sec:afterglow}), 
then the GRB cosmic-ray production rate is [see Eq.
(\ref{eq:gamma_rate})]
\begin{equation}
\epsilon_p^2 (d\dot n_p/d\epsilon_p)_{z=0}\approx 
10^{44}{\rm erg/Mpc}^3{\rm yr}.
\label{eq:cr_rate}
\end{equation}
\begin{figure}
\begin{center}
\includegraphics[width=3.5in]{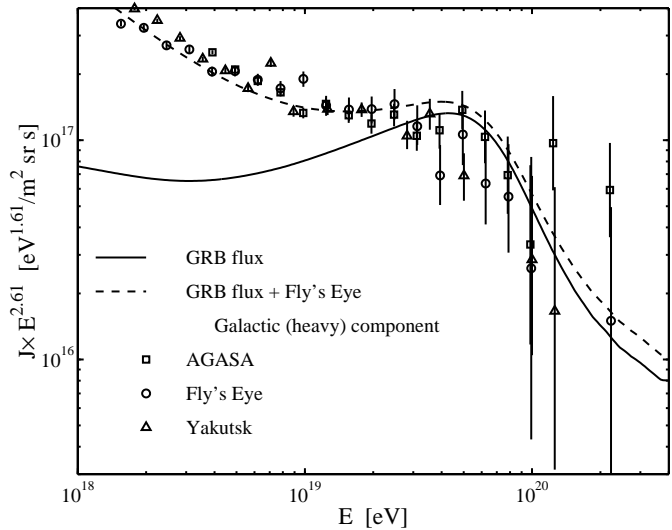}
\end{center}
\caption{
The UHECR flux expected in a cosmological model, where high-energy protons 
are produced at a rate $(\epsilon_p^2 d\dot n_p/d\epsilon_p)_{z=0}=
0.8\times10^{44}{\rm erg/Mpc}^3{\rm yr}$ as predicted by the GRB model 
[Eq. (\ref{eq:cr_rate})], solid line, compared to the Fly's Eye \cite{Bird94}, 
Yakutsk \cite{Yakutsk} and AGASA \cite{agasa} data. 
$1\sigma$ flux error bars are shown. The highest energy points are derived
assuming the detected events represent a
uniform flux over the energy range $10^{20}$~eV--$3\times10^{20}$~eV.
The dashed line is the sum of the GRB model flux and the Fly's Eye fit to
the Galactic heavy nuclei component, $J\propto \epsilon^{-3.5}$ \cite{Bird94}
(with normalization increased by 25\%). }
\label{fig1}
\end{figure}

In Fig. 1 we compare the UHECR spectrum,
reported by the Fly's Eye \cite{Bird94}, the Yakutsk \cite{Yakutsk}, 
and the AGASA experiments \cite{agasa}, 
with that expected from 
a homogeneous cosmological distribution of sources, each generating
a power law differential spectrum of high energy protons
$dn/d\epsilon_p\propto \epsilon_p^{-2}$. The absolute flux measured at
$3\times10^{18}$~eV differs between the various experiments,
corresponding to a systematic $\simeq10\%$ ($\simeq20\%$) over-estimate 
of event energies in the AGASA (Yakutsk)
experiment compared to the Fly's Eye experiment (see also \cite{Yoshida95}).
In Fig. 1, the Yakutsk energy normalization is used. 
For the model calculation, a flat universe, $\Omega=0.3$, $\Lambda=0.7$ and
$H_0=65{\rm km/ Mpc\ s}$ were assumed. The calculation is similar
to that described in \cite{W95b}. The generation rate of cosmic-rays 
(per unit comoving volume)
was assumed to evolve rapidly with redshift following
the luminosity density evolution of QSOs \cite{QSO}, which is 
also similar to that describing the evolution of star formation rate 
\cite{SFR1,SFR2}: $\dot n_{CR}(z)\propto(1+z)^{\alpha}$ with $\alpha\approx3$ 
\cite{QSOl} at low redshift, 
$z<1.9$, $\dot n_{CR}(z)={\rm Const.}$ for $1.9<z<2.7$, 
and an exponential decay at 
$z>2.7$ \cite{QSOh}. The cosmic-ray spectrum at energy $>10^{19}$~eV is 
little affected by modifications of the cosmological parameters or of
the redshift evolution of cosmic-ray generation rate. This is due to the
fact that cosmic-rays at this energy originate from distances shorter than
several hundred Mpc. The spectrum and flux at $\epsilon_p>10^{19}$~eV 
is mainly 
determined by the present ($z=0$) generation rate and spectrum, which
in the model shown in Fig. 1 is $\epsilon_p^2 
(d\dot n_p/d\epsilon)_{z=0}=0.8\times10^{44}
{\rm erg/Mpc}^3{\rm yr}$.

The suppression of model flux above $10^{19.7}$~eV is  
due to energy loss of high energy protons
in interaction with the microwave background, i.e. to the ``GZK cutoff''
\cite{GZK1,GZK2}. The available data do not allow to determine
the existence (or absence) of the ``cutoff''
with high confidence. The AGASA results show an excess (at a
$\sim2.5\sigma$ confidence level) of events compared to model
predictions above $10^{20}{\rm eV}$. This excess is not confirmed,
however, by the other experiments. Moreover, since the $10^{20}{\rm eV}$
flux is dominated by sources at distances $<100\ {\rm Mpc}$, over
which the distribution of known astrophysical systems
(e.g. galaxies, clusters of galaxies) is inhomogeneous,
 significant deviations from model predictions presented
in Fig. 1 for a uniform source distribution are expected at this energy
\cite{W95b}.
Clustering of cosmic-ray sources leads
to a standard deviation, $\sigma$, in the expected number, $N$, of 
events above $10^{20}$ eV, given by 
$\sigma /N = 0.9(d_0/10 {\rm Mpc})^{0.9}$
\cite{CR_clustering}, where $d_0$ is the unknown scale
length of the source correlation function and $d_0\sim10$ Mpc 
for field galaxies.

Thus, GRB fireballs would
produce UHECR flux and spectrum 
consistent with that observed, provided the efficiency
with which the wind kinetic energy is converted to $\gamma$-rays, and
therefore to electron energy, is similar to the efficiency with which it is
converted to proton energy, i.e. to UHECRs \cite{W95a}.
There is, however, one additional point which requires
consideration \cite{W95a}. The energy of the most
energetic cosmic ray detected by the Fly's Eye experiment is in excess of
$2\times10^{20}{\rm eV}$, and that of the most
energetic AGASA event is $\sim2\times10^{20}{\rm eV}$. On a
cosmological scale, the distance traveled by such energetic particles is
small: $<100{\rm Mpc}$ ($50{\rm Mpc}$) for the AGASA (Fly's Eye) event
(e.g., \cite{Aharonian94}). Thus, the detection of these events over a $\sim5
{\rm yr}$ period can be reconciled with the rate of nearby GRBs, $\sim1$
per $100\, {\rm yr}$ to $\sim1$
per $1000\, {\rm yr}$ out to $100{\rm Mpc}$, only if
there is a large dispersion, $\geq100{\rm yr}$, in the arrival time of protons 
produced in a single burst (This implies that if a direct 
correlation between 
high energy CR events and GRBs, as suggested in
\cite{MnU95}, is observed
on a $\sim10{\rm yr}$ time scale, it would be strong evidence {\it against} a 
cosmological GRB origin of UHECRs). 

The required dispersion
is likely to occur due to the combined effects of deflection 
by random magnetic fields and energy dispersion of the particles
\cite{W95a}. 
Consider a proton of energy $\epsilon_p$ 
propagating through a magnetic field of 
strength $B$ and correlation length
$\lambda$. As it travels a distance $\lambda$, the proton is typically 
deflected by an angle $\alpha\sim\lambda/
R_L$, where $R_L=\epsilon_p/eB$ is the Larmor radius. The
typical deflection angle for propagation over a distance $D$ is
$\theta_s\approx(2D/9\lambda)^{1/2}\lambda/R_L$. 
This deflection results in a time
delay, compared to propagation along a straight line, 
\begin{equation}
\tau(\epsilon_p,D)\approx\theta_s^2D/4c\approx
10^7\epsilon^{-2}_{p,20}D_{100}^2
\lambda_{\rm Mpc}B_{-8}^2\quad{\rm yr},
\label{delay}
\end{equation}
where $D=100D_{100}{\rm Mpc}$, $\lambda=1\lambda_{\rm Mpc}$~Mpc
and $B=10^{-8}B_{-8}$~G. Here, we have chosen numerical values 
corresponding to the current upper bound on the inter-galactic magnetic 
field, $B\lambda^{1/2}\le10^{-8}{\rm G\ Mpc}^{1/2}$ 
\cite{Kronberg77,Kronberg94}. The upper bound on the (systematic
increase with redshift of the) Faraday rotation 
measure of distant, $z\le2.5$, radio sources, $RM<5{\rm rad/m}^2$,
implies an upper bound $B\le10^{-11}(h/0.75)(\Omega_b h^2)^{-1}$~G
on an inter-galactic field coherent over cosmological scales \cite{Kronberg94}.
Here, $h$ is the Hubble constant in units of $100{\rm km/s\,Mpc}$ and
$\Omega_b$ is the baryon density in units of the closure density. For a
magnetic field coherent on scales $\sim\lambda$, this implies
$B\lambda^{1/2}\le10^{-8}(h/0.65)^{1/2}
(\Omega_b h^2/0.04)^{-1}{\rm G\ Mpc}^{1/2}$.

The random energy loss UHECRs suffer as they propagate, owing to the 
production of pions, implies that 
at any distance from the observer there is some finite spread
in the energies of UHECRs that are observed with a given fixed energy.
For protons with energies
$>10^{20}{\rm eV}$ the fractional RMS energy spread is of order unity
over propagation distances in the range $10-100{\rm Mpc}$ 
(e.g. \cite{Aharonian94}).
Since the time delay is sensitive to the particle energy, this implies that
the spread in arrival time of UHECRs with given observed energy is comparable
to the average time delay at that energy $\tau(\epsilon_p,D)$
(This result has been confirmed by numerical calculations in \cite{Coppi96}).
Thus, the  required time spread, $\tau>100$~yr, is consistent with
the upper bound, $\tau<10^7$~yr, implied by the present
upper bound to the inter-galactic magnetic field.

\section{GRB model predictions for UHECR experiments}
\label{sec:CRpred}

\subsection{The Number and Spectra of Bright Sources}
\label{sec:source-number}

The initial proton energy, necessary to have an observed energy $\epsilon_p$,
increases with source distance due to propagation energy losses.
The rapid increase of the initial energy after it exceeds, due to
electron-positron production, the threshold for pion production effectively
introduces a cutoff distance, $D_c(\epsilon_p)$, 
beyond which sources do not contribute
to the flux above $\epsilon_p$. 
The function $D_c(\epsilon_p)$ is shown in Fig. \ref{figNc} (adapted from
\cite{MnW96}). Since $D_c(\epsilon_p)$ is a decreasing function of 
$\epsilon_p$, for
a given number density of sources 
there is a critical energy $\epsilon_c$, above which
only one source (on average) contributes to the flux. 
In the GRB model $\epsilon_c$ depends on the product of the burst 
rate $R_{GRB}$
and the time delay. The number of sources contributing, on average, 
to the flux at energy $\epsilon_p$ is \cite{MnW96}
\begin{equation}
N(\epsilon_p) = {4\pi\over 5} R_{GRB}D_c(\epsilon_p)^3 
\tau\left[\epsilon_p,D_c(\epsilon_p)\right]\quad,
\label{N}
\end{equation}
and the average intensity resulting from all sources is
\begin{equation}
J(\epsilon_p) = \frac{1}{4\pi}R_{GRB} {d n_p\over d\epsilon_p} 
D_c(\epsilon_p)\quad,
\label{J}
\end{equation}
where $d n_p/d\epsilon_p$ 
is the number per unit energy of protons produced on average
by a single burst (this is the formal definition of $D_c(\epsilon_p)$). 
The critical energy $\epsilon_c$ is given by
\begin{equation}
{4\pi\over 5} R_{GRB}D_c(\epsilon_c)^3 \tau\left[\epsilon_c,
D_c(\epsilon_c)\right]=1\quad.
\label{Ec}
\end{equation}

\begin{figure}
\begin{center}
\includegraphics[width=3.5in]{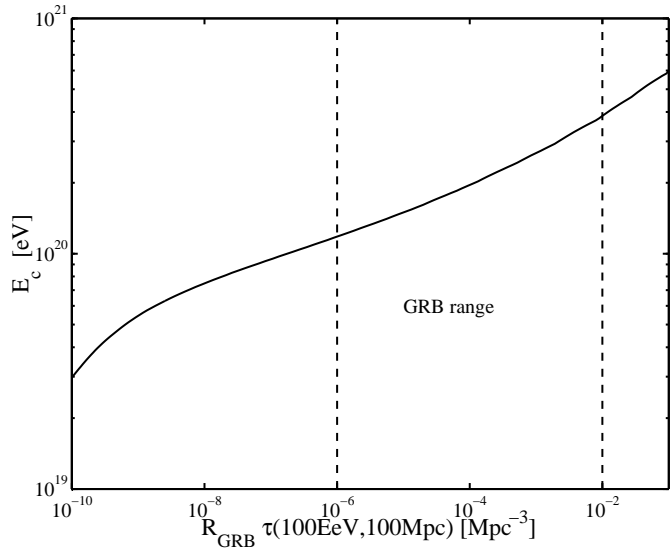}
\end{center}
\caption{$\epsilon_c$, the energy beyond which 
a single GRB contributes on average to
the UHECR flux, as a function of the product of GRB rate,
$R_{\rm GRB}\approx1/{\rm Gpc}^3$, and the time delay of a $10^{20}$~eV
proton originating at $100$~Mpc distance. The time delay depends
on the unknown inter-galactic field, $\tau\propto B^2\lambda$. Dashed
lines show the allowed range of $B^2\lambda$:
The lower limit is set by the requirement that 
at least a few GRB sources be present at $D<100$~Mpc, and the upper limit by 
the Faraday rotation bound 
$B\lambda^{1/2}\le10^{-8}{\rm G\ Mpc}^{1/2}$ \cite{Kronberg94}, see
Eq. (\ref{delay}) and the discussion the follows it.
}
\label{fig:Ec}
\end{figure}

\begin{figure}
\begin{center}
\includegraphics[width=3.5in]{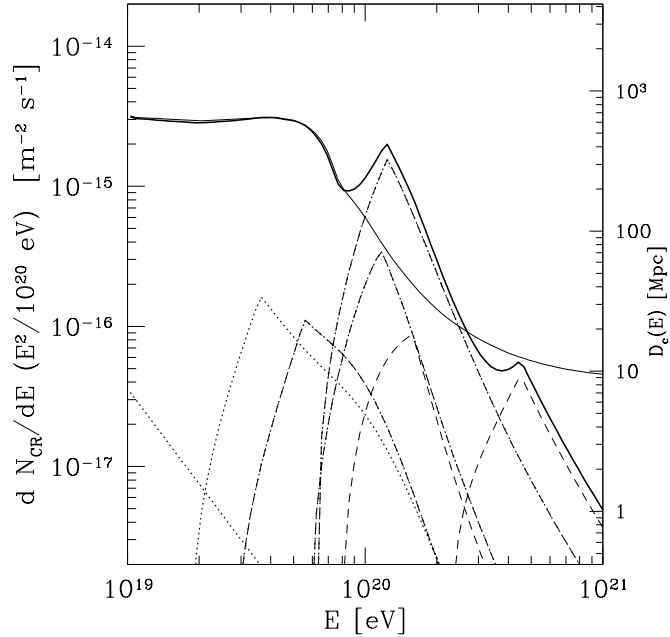}
\end{center}
\caption{Results of a Monte-Carlo realization of the bursting sources
model, with $\epsilon_c=1.4\times10^{20}$~eV: Thick solid line- overall 
spectrum in the realization;
Thin solid line- average spectrum, this
curve also gives $D_c(\epsilon_p)$;
Dotted lines- spectra of brightest sources at different energies.
}
\label{figNc}
\end{figure}

$\epsilon_c$, the energy beyond which 
a single source contributes on average to
the flux, depends on the unknown properties of 
the inter-galactic magnetic field, $\tau\propto B^2\lambda$. 
However, the rapid
decrease of $D_c(\epsilon_p)$ with energy near $10^{20}{\rm eV}$
implies that $\epsilon_c$ is only weakly dependent 
on the value of $B^2\lambda$, as shown in Fig. \ref{fig:Ec}. 
In The GRB model, the product $R_{GRB}\tau(D=100{\rm Mpc},\epsilon_p
=10^{20}{\rm eV})$
is approximately limited to the range $10^{-6}{\rm\ Mpc}^{-3}$ to
$10^{-2}{\rm\ Mpc}^{-3}$ [The lower limit is set by the requirement that 
at least a few GRB sources be present at $D<100$~Mpc, and the upper limit by 
the Faraday rotation bound 
$B\lambda^{1/2}\le10^{-8}{\rm G\ Mpc}^{1/2}$ \cite{Kronberg94}, see
Eq. (\ref{delay}), and 
$R_{GRB}\le1/{\rm\ Gpc}^3{\rm yr}$]. The corresponding range
of values of $\epsilon_c$ is 
$10^{20}{\rm eV}\le \epsilon_c<4\times10^{20}{\rm eV}$.

Fig. \ref{figNc} presents the flux obtained in one realization of
a Monte-Carlo simulation described by Miralda-Escud\'e \& Waxman 
\cite{MnW96} of the total
number of UHECRs received from GRBs at some fixed time. 
For each
realization the positions (distances from Earth) and
times at which cosmological GRBs occurred were randomly drawn, 
assuming an intrinsic proton
generation spectrum $dn_p/d\epsilon_p \propto \epsilon_p^{-2}$, and 
$\epsilon_c=1.4\times10^{20}{\rm eV}$. 
Most of the realizations gave an overall spectrum similar to that obtained
in the realization of Fig. \ref{figNc} when the brightest source of this 
realization (dominating at $10^{20}{\rm eV}$) is not included.
At $\epsilon_p < \epsilon_c$,
the number of sources contributing to the flux is very large, 
and the overall UHECR flux received at any
given time is near the average (the average flux is that obtained when 
the UHECR emissivity is spatially uniform and time independent).
At $\epsilon_p > \epsilon_c$, the flux will generally be much lower than the average,
because there will be no burst within a distance $D_c(\epsilon_p)$ having taken
place sufficiently recently. There is, however, a significant probability
to observe one source with a flux higher than the average.
A source similar to the brightest one in Fig. \ref{figNc}
appears $\sim5\%$ of the time. 

At any fixed time a given burst is observed in UHECRs only over a narrow
range of energy, because if
a burst is currently observed at some energy $\epsilon_p$ 
then UHECRs of much lower 
energy from this burst have not yet arrived,  while higher energy UHECRs
reached us mostly in the past. As mentioned above, for energies above the 
pion production threshold, 
$\epsilon_p\sim5\times10^{19}{\rm eV}$, the dispersion in arrival times of UHECRs
with fixed observed energy is comparable to the average delay at that
energy. This implies that
the spectral width $\Delta \epsilon_p$ of the source at a given time is of order
the average observed energy, $\Delta \epsilon_p\sim \epsilon_p$.
Thus, bursting UHECR sources should have narrowly peaked energy
spectra,
and the brightest sources should be different at different energies.
For steady state sources, on the other hand, the brightest
source at high energies should also be the brightest one at low
energies, its fractional contribution to the overall flux decreasing to
low energy only as $D_c(\epsilon_p)^{-1}$.
A detailed numerical analysis of the time dependent energy spectrum of 
bursting sources is given in \cite{Sigl97,Lemoine97}.

\subsection{Spectra of Sources at $\epsilon_p<4\times10^{19}{\rm eV}$}
\label{subsec:Blambda}

The detection of UHECRs 
above $10^{20}{\rm eV}$ imply that the brightest sources 
must lie at distances smaller than $100{\rm Mpc}$.
UHECRs with $\epsilon_p\le4\times10^{19}{\rm eV}$
from such bright sources will suffer energy loss only by pair production,
because at $\epsilon_p < 5\times 10^{19}$ eV
the mean-free-path for pion production interaction
(in which the fractional energy loss is $\sim10\%$) is larger than 
$1{\rm Gpc}$. Furthermore, the energy loss due to pair production 
over $100{\rm Mpc}$ propagation is only $\sim5\%$.

In the case where the typical displacement of the UHECRs 
due to deflections by inter-galactic magnetic fields is 
much smaller than the correlation length, $\lambda \gg D\theta_s(D,\epsilon_p)
\simeq D(D/\lambda)^{1/2}\lambda/R_L$,
all the UHECRs that arrive at the
observer are essentially deflected by the same magnetic field structures, 
and the absence of random energy loss during propagation implies that
all rays with a fixed observed energy would reach the observer with exactly
the same direction and time delay. At a fixed time, therefore, the source would
appear mono-energetic and point-like. In reality,
energy loss due to pair production
results in a finite but small spectral and angular width, 
$\Delta \epsilon_p/\epsilon_p\sim\delta\theta/\theta_s\le1\%$ \cite{WnM96}.

In the case where the typical displacement of the UHECRs is 
much larger than the correlation length, $\lambda \ll D\theta_s(D,\epsilon_p)$,
the deflection of different UHECRs arriving at the observer
are essentially independent. Even in the absence of any energy loss there 
are many paths from the source to the observer for UHECRs of fixed energy $\epsilon_p$
that are emitted from the source at an angle 
$\theta\le\theta_s$ relative to the source-observer line of sight. Along
each of the paths, UHECRs are deflected by independent magnetic field 
structures. Thus, the source angular size would be of order $\theta_s$
and the spread in arrival times would be comparable to the characteristic 
delay $\tau$, leading to $\Delta \epsilon_p/\epsilon_p\sim1$ even when there are no random
energy losses. The observed spectral shape of a nearby ($D<100{\rm Mpc}$) 
bursting source of UHECRs at 
$\epsilon_p<4\times10^{19}{\rm eV}$
was derived for the case $\lambda \ll D\theta_s(D,\epsilon_p)$ in 
\cite{WnM96}, and is given by
\begin{equation}
{dN\over d\epsilon_p}\propto \sum\limits_{n=1}^{\infty} (-1)^{n+1}\, n^2\,
\exp\left[ -{2n^2\pi^2 \epsilon^2\over \epsilon_0^2(t,D)} \right]\quad,
\label{flux}
\end{equation}
where $\epsilon_0(t,D)=De(2{B^2\lambda}/3ct)^{1/2}$.  
For this spectrum, the ratio of the 
RMS UHECR energy spread to the average energy is $30\%$

\begin{figure}
\begin{center}
\includegraphics[width=3.5in]{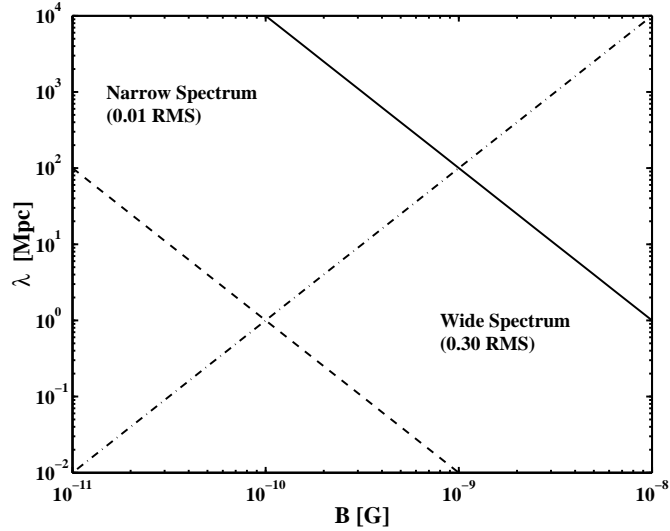}
\end{center}
\caption{The line $\theta_s D=\lambda$ for a source at 
$30{\rm Mpc}$ distance 
observed at energy $\epsilon_p\simeq10^{19}{\rm eV}$ (dot-dash line), 
shown with
the Faraday rotation upper limit  $B\lambda^{1/2}
\le10^{-8}{\rm G\ Mpc}^{1/2}$ (solid line), and with the lower limit 
$B\lambda^{1/2}\ge10^{-10}{\rm G\ Mpc}^{1/2}$ required in the GRB model
[see Eq. (\ref{delay})].
}
\label{figBL}
\end{figure}

Fig. 
\ref{figBL} shows the line $\theta_s D=\lambda$ in the $B-\lambda$ plane,
for a source at a distance $D=30{\rm Mpc}$
observed at energy $\epsilon_p\simeq10^{19}{\rm eV}$.
Since the $\theta_s D=\lambda$ line divides the
allowed region in the plane at $\lambda\sim1{\rm Mpc}$,
measuring the spectral width of bright sources would allow to determine
if the field correlation length is much larger, much smaller, or comparable
to $1{\rm Mpc}$.

\section{High energy Neutrinos}
\label{sec:nus}

\subsection{Internal shock (GRB) neutrinos}
\label{sec:GRBnu}

\subsubsection{Neutrinos at energies $\sim10^{14}$~eV.}

Protons accelerated in the fireball to high energy lose energy through
photo-meson interaction with fireball photons. The decay of charged
pions produced in this interaction, $\pi^+\rightarrow\mu^++\nu_\mu
\rightarrow e^++\nu_e+\overline\nu_\mu+\nu_\mu$, 
results in the production of high energy neutrinos. 
The key relation is between the observed photon energy, $\epsilon_\gamma$,
and the accelerated proton's energy, $\epsilon_p$,
at the threshold of the $\Delta$-resonance.
In the observer frame,
\begin{equation}
\epsilon_\gamma \, \epsilon_{p} = 0.2 \, {\rm GeV^2} \, \Gamma^2\,.
\label{eq:keyrelation}
\end{equation}
For $\Gamma\approx300$ and $\epsilon_\gamma=1$~MeV,
we see that characteristic proton energies
$\sim 10^{16}$~eV
are required to produce pions. Since neutrinos produced by pion decay
typically carry $5\%$ of the proton energy (see below), production of
$\sim 10^{14}$~eV neutrinos is expected \cite{WnB97,WnB99}.

The fractional energy loss rate of
a proton with energy $\epsilon'_p=\epsilon_p/\Gamma$ 
measured in the wind rest frame due to
pion production is
\begin{eqnarray}
t_\pi^{-1}(\epsilon'_p)&&\equiv
-{1\over\epsilon'_p}{d\epsilon'_p\over dt}\nonumber\\
&&={1\over2\gamma_p^2}c\int_{\epsilon_0}^\infty 
d\epsilon\,\sigma_\pi(\epsilon)
\xi(\epsilon)\epsilon\,\int_{\epsilon/2\Gamma_p}^\infty dx\,
x^{-2}\frac{dn_\gamma}{d\epsilon_\gamma}(\epsilon_\gamma=x)\,,
\label{eq:pirate}
\end{eqnarray}
where $\gamma_p=\epsilon'_p/m_pc^2$, $\sigma_\pi(\epsilon)$ is the cross
section for pion production for a photon with energy $\epsilon$ in the
proton rest frame, $\xi(\epsilon)$ is the average fraction of energy
lost to the pion, $\epsilon_0=0.15{\rm GeV}$ is the threshold
energy, and $dn_\gamma/d\epsilon_\gamma$ is the photon density per unit
photon energy in the wind rest frame. In deriving Eq. (\ref{eq:pirate})
we have assumed that the photon distribution in the wind rest frame is 
isotropic. The GRB photon spectrum is well fitted in the
BATSE detector range, 20~keV to 2~MeV, by a combination
of two power-laws, 
$dn_\gamma/d\epsilon_\gamma\propto\epsilon_\gamma^{-\beta}$,
with $\beta\simeq1$ at $\epsilon_\gamma<\epsilon_{\gamma b}$,
$\beta\simeq2$ at $\epsilon_\gamma>\epsilon_{\gamma b}$ and 
$\epsilon_{\gamma b}\sim1{\rm MeV}$ \cite{Band}. Thus, 
the second integral in (\ref{eq:pirate}) may be approximated by
\begin{equation}
\int_\epsilon^\infty{\rm d}x\,x^{-2}
\frac{dn_\gamma}{d\epsilon_\gamma}(\epsilon_\gamma=x)\simeq
{1\over1+\beta}{U_\gamma\over2\epsilon^{\prime 3}_{\gamma b}}
\left({\epsilon\over\epsilon'_{\gamma b}}\right)^{-(1+\beta)}\,,
\end{equation}
where $U_\gamma$ is the photon energy density (in the range
corresponding to the observed BATSE range) in the wind rest-frame,
$\beta=1$ for $\epsilon<\epsilon'_{\gamma b}$ and 
$\beta=2$ for $\epsilon>\epsilon'_{\gamma b}$.
$\epsilon'_{\gamma b}$ is the break energy measured in the wind frame,
$\epsilon'_{\gamma b}=\epsilon_{\gamma b}/\Gamma$.
The main contribution to the first integral in (\ref{eq:pirate}) is from 
photon energies $\epsilon\sim\epsilon_{\rm peak}=
0.3{\rm GeV}$, where the cross section
peaks due to the $\Delta$ resonance. Approximating the integral by the
contribution from the resonance we obtain
\begin{equation}
t_\pi^{-1}(\epsilon'_p)\simeq
{U_\gamma\over2\epsilon'_{\gamma b}}c\sigma_{\rm peak}\xi_{\rm peak}
{\Delta\epsilon\over\epsilon_{\rm peak}}
\min(1,2\gamma_p\epsilon'_{\gamma b}/\epsilon_{\rm peak})\,.
\label{taupi}
\end{equation}
Here, $\sigma_{\rm peak}\simeq5\times10^{-28}{\rm cm}^2$ and
$\xi_{\rm peak}\simeq0.2$ are the values of $\sigma$ and $\xi$
at $\epsilon=\epsilon_{\rm peak}$, and $\Delta\epsilon\simeq
0.2{\rm GeV}$ is the peak width.

The energy loss of protons due to pion production is small during the
acceleration process. Once accelerated, the time available
for proton energy loss by pion production is comparable to
the wind expansion time as measured in the wind rest frame, $t_{\rm co}\sim r/
\Gamma c$. Thus, the fraction of energy lost by protons to pions is 
$f_\pi\simeq r/\Gamma ct_\pi$. 
The energy density in the BATSE range, $U_\gamma$, is related to the
luminosity $L_\gamma$ by $L_\gamma=4\pi r^2\Gamma^2cU_\gamma$. 
Using this relation in (\ref{taupi}), and $r=2\Gamma^2 c\Delta t$,
$f_\pi$ is given by \cite{WnB97}
\begin{eqnarray}
f_\pi(\epsilon_p)\approx0.1{L_{\gamma,52}\over
\epsilon_{\gamma b,{\rm MeV}}\Gamma_{2.5}^4 \Delta t_{-2}}
\times
\cases{1,&if $\epsilon_{p}>\epsilon_{pb}$;\cr
\epsilon_{p}/\epsilon_{pb},&otherwise.\cr}
\label{eq:fpi}
\end{eqnarray}
The proton break energy is
\begin{equation}
\epsilon_{pb}\approx10^{16}\Gamma_{2.5}^2
(\epsilon_{\gamma b,{\rm MeV}})^{-1}\, 
{\rm eV}\,.
\label{epb}
\end{equation}

The value of $f_\pi$, Eq. (\ref{eq:fpi}), is strongly dependent on 
$\Gamma$. It has recently been pointed out \cite{HnH99}
that if the Lorentz factor $\Gamma$ varies significantly
between bursts, with burst to burst variations $\Delta\Gamma/\Gamma\sim1$, 
then the resulting neutrino
flux will be dominated by a few neutrino bright bursts, and may 
significantly exceed 
the flux implied by Eq. (\ref{eq:fpi}), derived 
for typical burst parameters.
This may strongly enhance the detectability of GRB neutrinos
by planned neutrino telescopes \cite{Alvarez00}.
However, as explained in \S\ref{sec:fireball-gammas}, the Lorentz factors
of fireballs producing observed GRBs can not differ significantly from
the minimum value allowed by Eq. (\ref{eq:Gmingg}), $\Gamma\simeq250$, 
for which the fireball pair production optical depth, Eq. (\ref{eq:taugg}),
is $\approx1$ for $\epsilon_\gamma=100$~MeV: 
Lower Lorentz factors lead to optically thick fireballs,
while higher Lorentz factors lead to low luminosity X-ray bursts (which may
have already been identified). Thus, for Lorentz factors consistent
with observed GRB spectra, for which 
$\tau_{\gamma\gamma}(\epsilon_\gamma=100{\rm MeV})\approx1)$, we find
\begin{eqnarray}
f_\pi(\epsilon_p)\approx0.2{L_{\gamma,52}^{1/3}\over
\epsilon_{\gamma b,{\rm MeV}}\Delta t_{-2}^{1/3}}
\times
\cases{1,&if $\epsilon_{p}>\epsilon_{pb}$;\cr
\epsilon_{p}/\epsilon_{pb},&otherwise.\cr}
\label{eq:fpi1}
\end{eqnarray}
A detailed analysis, using Monte-Carlo simulations of the internal shock
model, confirms that for fireball parameter range consistent with observed
GRB characteristics, $f_\pi$ at $\epsilon_{p}>\epsilon_{pb}$
is limited to the range of $\sim10\%$ to $30\%$ \cite{GSW01}.

Thus, for parameters typical of a GRB producing wind, a significant fraction
of the energy of protons accelerated to energies larger than
the break energy, $\sim10^{16}{\rm eV}$, would be lost to pion production.
Roughly half of the energy lost by protons goes into $\pi^0$~'s and the 
other half to $\pi^+$~'s.  Neutrinos are produced by the decay of $\pi^+$'s, 
$\pi^+\rightarrow\mu^++\nu_\mu
\rightarrow e^++\nu_e+\overline\nu_\mu+\nu_\mu$ [the large optical
depth for high energy $\gamma$'s from $\pi^0$ decay, Eq. (\ref{eq:taugg}),
would not allow these photons to escape the wind]. 
The mean pion energy is $20\%$ 
of the energy of the proton producing the pion. This energy is roughly
evenly distributed between 
the $\pi^+$ decay products. Thus, approximately half 
the energy lost by protons of energy $\epsilon_p$ is converted to neutrinos 
with energy $\sim0.05\epsilon_p$. Eq. (\ref{eq:fpi}) then implies that 
the spectrum of neutrinos above $\epsilon_{\nu b}=0.05\epsilon_{pb}$
follows the proton spectrum, and is harder
(by one power of the energy) at lower energy.

If GRBs are the sources of UHECRS, 
then using Eq. (\ref{eq:cr_rate})
the expected GRB neutrino flux is \cite{WnB99}
\begin{eqnarray}
\epsilon_\nu^2\Phi_{\nu_\mu}&&\approx \epsilon_\nu^2\Phi_{\bar\nu_\mu}
\approx \epsilon_\nu^2\Phi_{\nu_e}
\approx\frac{c}{4\pi}\frac{f_\pi}{8}\epsilon_p^2 (d\dot n_p/d\epsilon_p)t_H
\cr&&\approx 1.5\times10^{-9}{f_\pi(\epsilon_{p b})\over0.2}
\min\{1,\epsilon_\nu/\epsilon_{\nu b}\}
{\rm GeV\,cm}^{-2}{\rm s}^{-1}{\rm sr}^{-1},
\label{eq:JGRB}
\end{eqnarray}
where $t_H\approx10^{10}$~yr is the Hubble time. The factor of $1/8$ is due to
the fact that charged pions and neutral pions are produced with
roughly equal probabilities (and each neutrino carries $\sim1/4$ of the
pion energy).

The GRB neutrino flux can also be estimated directly from
the observed gamma-ray fluence. The  
BATSE detectors 
measure the GRB fluence $F_\gamma$ over two decades of photon energy, 
$\sim0.02$MeV to $\sim2$MeV, corresponding to a decade of radiating
electron energy (the electron 
synchrotron frequency is proportional to the square of
the electron Lorentz factor). If electrons carry a fraction $\xi_e$ of
the energy carried by protons, then the muon neutrino fluence of a single
burst is 
$\epsilon_\nu^2dN_\nu/d\epsilon_\nu\approx0.25(f_\pi/\xi_e)F_\gamma/\ln(10)$. 
The average neutrino flux per unit time and solid angle is obtained by
multiplying the single burst fluence with the GRB rate per solid angle,
$\approx10^3$ bursts per year over $4\pi$~sr. Using the average burst
fluence $F_\gamma=10^{-5}{\rm erg/cm}^2$, we obtain
a muon neutrino flux $\epsilon_\nu^2\Phi_\nu\approx3\times10^{-9}(f_\pi/\xi_e)
{\rm GeV/cm}^2{\rm s\,sr}$. Thus, the 
neutrino flux estimated directly from the gamma-ray fluence agrees with
the estimate (\ref{eq:JGRB}) based on the cosmic-ray production rate.

\subsubsection{Neutrinos at energy $>10^{16}$~eV.}

The neutrino spectrum (\ref{eq:JGRB}) is
modified at high energy, where neutrinos are produced by the decay
of muons and pions whose life time $\tau_{\mu,\pi}$
exceeds the characteristic time for
energy loss due to adiabatic expansion and synchrotron emission 
\cite{WnB97,RnM98,WnB99}.
The synchrotron loss time is determined by the energy density of the
magnetic field in the wind rest frame.
For the characteristic parameters of a GRB wind, the muon energy for which
the adiabatic energy loss time equals the muon life time,
$\epsilon^a_\mu$, is comparable to the energy $\epsilon^s_\mu$ 
at which the life time equals 
the synchrotron loss time, $\tau^s_\mu$. For pions, $\epsilon^a_\pi>\epsilon^s_\pi$. This,
and the fact that the adiabatic
loss time is independent of energy and the synchrotron loss time is
inversely proportional to energy, imply that
synchrotron losses are the dominant effect
suppressing the flux at high energy. The energy above which synchrotron
losses suppress the neutrino flux is
\begin{equation}
{\epsilon^s_{\nu_\mu(\bar\nu_\mu,\nu_e)}\over \epsilon_{\nu b}}
\approx\left(\frac{\xi_B}{\xi_e} L_{\gamma,52}\right)^{-1/2}\Gamma_{2.5}^2
\Delta t_{-2}\epsilon_{\gamma b,\rm MeV}\times
\cases{10,&for $\bar\nu_\mu$, $\nu_e$;\cr 100,&for $\nu_\mu$ .\cr}
\label{eq:nu_sync}
\end{equation}

The efficiency of neutrino production in internal collisions 
decreases with $\Delta t$,
$f_\pi\propto\Delta t^{-1}$ [see Eq. (\ref{eq:fpi})], 
since the radiation energy 
density is lower at larger collision radii. However, at larger radii 
synchrotron losses cut off
the spectrum at higher energy, $\epsilon^s(\Delta t)\propto\Delta t$ [see Eq. 
(\ref{eq:nu_sync})]. Collisions at large radii therefore
result in extension of the neutrino
spectrum of Eq. (\ref{eq:JGRB}) to higher energy, beyond the cutoff energy 
Eq. (\ref{eq:nu_sync}), 
\begin{equation}
\epsilon^2_\nu\Phi_\nu\propto 
\epsilon_\nu^{-1},\quad \epsilon_\nu>\epsilon^s_\nu.
\label{high_nu}
\end{equation}

\subsubsection{Comparison with other authors.}

We note, that the results presented above were derived using the 
$\Delta$-approximation, i.e.
assuming that photo-meson interactions are dominated by the contribution of
the $\Delta$-resonance.
It has recently been shown \cite{Muecke98}, that for photon spectra harder
than $dn_\gamma/d\epsilon_\gamma\propto \epsilon^{-2}_\gamma$, 
the contribution of 
non-resonant interactions may be important. Since in order to interact with
the hard part of the photon spectrum, $\epsilon_\gamma<\epsilon_{\gamma b}$, 
the proton energy
must exceed the energy at which neutrinos of energy $\epsilon_{\nu b}$ are
produced, significant modification of the $\Delta$-approximation results
is expected only for $\epsilon_\nu\gg \epsilon_{\nu b}$, where the 
neutrino flux is 
strongly suppressed by synchrotron losses.

The neutrino flux from GRBs is small above $10^{19}$eV, 
and a neutrino flux comparable to the $\gamma$-ray flux
is expected only below $\sim10^{17}$eV, in agreement with the
results of ref. \cite{RnM98}.
Our result is not in agreement, however, with that of ref. \cite{Vietri_nu19}, 
where
a much higher flux at $\sim10^{19}$eV is obtained based on the equations
of ref. \cite{WnB97}, which are the same equations as used 
here\footnote{The parameters chosen in \cite{Vietri_nu19} 
are $L_\gamma=10^{50}{\rm
erg/s}$, $\Delta t=10$s, and $\Gamma=100$. Using equation (4) of ref.
\cite{WnB97}, which is the same as Eq. (\ref{eq:fpi}) of the present paper,
we obtain for these parameters $f_\pi=1.6\times10^{-4}$, while the author
of \cite{Vietri_nu19} obtains, using the same equation, $f_\pi=0.03$.}.

\subsection{Reverse shock (afterglow) neutrinos, $\sim10^{18}$~eV}
\label{sec:AGnus}

Let us now consider neutrino emission from photo-meson interactions of 
protons accelerated to high energies  
in the reverse shocks driven into the fireball ejecta at the
initial stage of interaction of the fireball with its surrounding gas, which
occurs on time scale $T\sim10$~s, comparable to the duration of the GRB 
itself (see \S\ref{sec:fireball-interaction}).
Optical--UV photons are radiated  by electrons accelerated in shocks
propagating backward into the ejecta
(see \S\ref{sec:fireball-afterglow}), and may interact with accelerated
protons. The interaction of these low energy, 10~eV--1~keV, photons 
and high energy protons produces a burst of duration $\sim T$ 
of ultra-high energy, $10^{17}$--$10^{19}$~eV, neutrinos [as indicated by
Eq. (\ref{eq:keyrelation})] via photo-meson interactions \cite{WnB-AG}.

Afterglows have been detected in several cases; reverse
shock emission has only been identified for GRB 990123 \cite{Akerlof99}.
Both the detections and the non-detections are consistent with shocks
occurring with typical model parameters \cite{Draine00,SP_0123,MR_0123},
suggesting that reverse shock emission may be common.
The predicted neutrino emission depends, however, upon parameters
of the surrounding medium that
can only be estimated once
more observations of the prompt optical afterglow emission are available.
We first consider the case where the 
density of gas surrounding the fireball is $n\sim1{\rm cm}^{-3}$,
a value typical to the inter-stellar medium and consistent with
GRB 990123 observations.

The photon density in Eq. (\ref{eq:pirate})
is related to the observed specific luminosity by
$dn_\gamma/d\epsilon_\gamma(x)=L_\epsilon(\Gamma x)/(4\pi r^2c\Gamma x)$.
For proton Lorentz factor
$\epsilon_0/2\epsilon'_{\gamma c}\ll\gamma_p<\epsilon_0/2\epsilon'_{\gamma m}$,
where primed energies denote rest frame energies 
(e.g. $\epsilon'_{\gamma m}=\epsilon_{\gamma m}/\Gamma_{\rm tr.}$),
photo-meson production is dominated by interaction with photons
in the energy range
$\epsilon_{\gamma m}<\epsilon_\gamma\ll\epsilon_{\gamma c}$,
where $L_\epsilon\propto\epsilon_\gamma^{-1/2}$
(see \S\ref{sec:fireball-afterglow}). For this photon spectrum,
the contribution to the first integral of Eq. (\ref{eq:pirate}) from
photons at the $\Delta$-resonance is comparable to that of photons of
higher energy, and we obtain
\begin{equation}
t_\pi^{-1}(\epsilon'_p)\simeq {2^{5/2}\over2.5}
{L_m\over4\pi r^2\Gamma_{\rm tr.}}
\left({\epsilon_{\rm peak}\over \gamma_p \epsilon'_{\gamma
m}}\right)^{-1/2}
 {\sigma_{\rm peak}\xi_{\rm peak}
\Delta\epsilon\over\epsilon_{\rm peak}}.
\label{eq:taupi}
\end{equation}
$\Gamma_{\rm tr.}$ is the expansion Lorentz factor of plasma shocked
by the reverse shocks, given by Eq. (\ref{eq:Gamma_tr}). The time available
for proton energy loss by pion production is comparable to
the expansion time as measured in the wind rest frame,
$\sim r/\Gamma_{\rm tr.} c$.
Thus, the fraction of energy lost by protons to pions is
\begin{eqnarray}
f_\pi(\epsilon_p)\approx 0.1
\left({L_m\over10^{61}{\rm s}^{-1}}\right)
\left({\Gamma_{\rm tr.}\over250}\right)^{-5} T_1^{-1}\times
(\epsilon_{\gamma m,{\rm eV}}\epsilon_{p,20})^{1/2}.
\label{eq:fpiAG}
\end{eqnarray}
Eq. (\ref{eq:fpiAG}) is valid for protons in the energy range
\begin{eqnarray}
4\times10^{18}\left({\Gamma_{\rm tr.}\over250}\right)^2
(\epsilon_{\gamma c,{\rm keV}})^{-1}{\rm eV}
<\epsilon_{p}<
4\times10^{21}\left({\Gamma_{\rm tr.}\over250}\right)^2
(\epsilon_{\gamma m,{\rm eV}})^{-1}{\rm eV}
\,.
\label{eq:ep_range}
\end{eqnarray}
Such protons interact with photons in the energy range $\epsilon_{\gamma m}$
to $\epsilon_{\gamma c}$, where the photon spectrum 
$L_\epsilon\propto\epsilon_{\gamma c}^{1/2}$ and the number of photons above
interaction threshold is $\propto\epsilon_p^{1/2}$.
At lower energy, protons interact with photons of energy
$\epsilon_\gamma>\epsilon_{\gamma c}$, where
$L_\epsilon\propto\epsilon^{-1}$ rather then
$L_\epsilon\propto\epsilon^{-1/2}$. At these energies therefore
$f_\pi\propto\epsilon_p$.

Since approximately half
the energy lost by protons of energy $\epsilon_p$ is converted to neutrinos
with energy $\sim0.05\epsilon_p$, Eq. (\ref{eq:ep_range}) implies that
the spectrum of neutrinos below
$\epsilon_{\nu b}\approx10^{17}(\Gamma_{\rm tr.}/250)^2
(\epsilon_{\gamma c,{\rm keV}})^{-1}{\rm eV}$ is harder
by one power of the energy then the proton spectrum, and
by half a power of the energy at higher energy. For a power law
differential spectrum of accelerated protons $dn_p/d\epsilon_p\propto
\epsilon_p^{-2}$, the differential neutrino spectrum is $dn_\nu/d\epsilon_\nu
\propto\epsilon_\nu^{-\alpha}$ with $\alpha=1$ below the break and
$\alpha=3/2$ above the break. Assuming that GRBs are 
indeed the sources of ultra-high energy cosmic rays,
then Eqs. (\ref{eq:fpiAG},\ref{eq:ep_range}) and (\ref{eq:cr_rate})
imply that the expected neutrino intensity is
\begin{eqnarray}
\epsilon_\nu^2\Phi_{\nu_\mu}&&\approx \epsilon_\nu^2\Phi_{\bar\nu_\mu}
\approx \epsilon_\nu^2\Phi_{\nu_e}\cr
&&\approx 
10^{-10}{f_\pi^{[19]}\over0.1}
\left({\epsilon_\nu\over10^{17}{\rm eV}}\right)^{\beta}
{\rm GeV\,cm}^{-2}{\rm s}^{-1}{\rm sr}^{-1},
\label{eq:JGRBAG}
\end{eqnarray}
where $f_\pi^{[19]}\equiv f_\pi(\epsilon_{p,20}=2)$ and
$\beta=1/2$ for $\epsilon_\nu>10^{17}{\rm eV}$
and $\beta=1$ for $\epsilon_\nu<10^{17}{\rm eV}$.

Some GRBs may result from the collapse of a massive star, in
which case the fireball is expected to expand into a pre-existing wind
(e.g. \cite{Chevalier00,WoosleyAG}). 
For typical wind parameters, the
transition to self-similar behavior takes place at a radius where the
wind density is $n\approx10^4{\rm cm}^{-3}\gg 1{\rm cm}^{-3}$. The higher
density implies a lower Lorenz factor of the expanding plasma during
the transition stage, and a larger fraction of proton energy lost
to pion production. Protons of energy
$\epsilon_p\ge 10^{18}$~eV lose all their energy to pion production
in this case. If most GRBs result from the collapse of massive stars, then
the predicted neutrino flux is \cite{WnB-AG,Dai00}
\begin{equation}
\epsilon_\nu^2\Phi_\nu\approx 10^{-8}\min\{1,
\epsilon_\nu^{\rm ob.}/10^{17}{\rm eV}\}
{\rm GeV\,cm}^{-2}{\rm s}^{-1}{\rm sr}^{-1}.
\label{eq:JGRBAGw}
\end{equation}

The neutrino flux is expected to be strongly suppressed at energy
$\epsilon_\nu>10^{19}$~eV, since protons are not expected to be
accelerated to energy $\epsilon_p\gg10^{20}$~eV.
If protons are accelerated to much higher energy, the $\nu_\mu$ 
($\bar\nu_\mu$, $\nu_e$) 
flux may extend to $\sim10^{21}n_0^{-1/2}\xi_{B,-1}^{-1/2}$~eV
($\sim10^{20}n_0^{-1/2}\xi_{B,-1}^{-1/2}$~eV). 
At higher energy, synchrotron losses of pions (muons) will suppress the
neutrino flux.

\subsection{Inelastic $p$-$n$ collisions}
\label{sec:pnnus}

The acceleration, $\gamma\propto r$, of fireball plasma emitted from the
source of radius $r_0$ (see \S\ref{sec:fireball-evolution}) 
is driven by radiation pressure. Fireball
protons are accelerated through their coupling to the electrons, which
are coupled to fireball photons. Fireball neutrons, which are expected to
exist in most progenitor scenarios, are coupled to protons by nuclear
scattering as long as the comoving $p$-$n$ scattering time
is shorter than the comoving wind expansion time $r/\gamma c=r_0/c$. 
As the fireball plasma expands and accelerates, the proton density decreases,
$n_p\propto r^{-2}\gamma^{-1}$, and neutrons may become decoupled.
For $\eta>\eta_{pn}$, where
\begin{equation}
\eta_{pn}\approx400L_{52}^{1/4}r_{0,7}^{-1/4},
\label{eq:Gamma_pn}
\end{equation}
neutrons decouple
from the accelerating plasma prior to saturation, $\gamma=\eta$,
at $\Gamma=\eta_{pn}^{4/3}\eta^{-1/3}$ \cite{Derishev99,BnM00}. In this
case, relativistic relative velocities between protons and neutrons arise,
which lead to pion production through inelastic nuclear collisions. Since
decoupling occurs at a radius where the collision time is similar to wind
expansion time, each $n$ leads on average to one pair of $\nu\bar\nu$.
The typical comoving neutrino energy, $\sim50$~MeV, implies an observed
energy $\sim10$~GeV. A typical burst, $E=10^{53}$~erg at $z=1$, with
significant neutron to proton ratio and $\eta>400$ will therefore
produce a fluence $F(\nu_e+\bar\nu_e)\sim0.5F(\nu_\mu+\bar\nu_\mu)\sim
10^{-4}{\rm cm}^{-2}$ of $\sim10$~GeV neutrinos.

Relativistic relative $p$-$n$ velocities, leading to neutrino
production through inelastic collisions, may also result from diffusion
of neutrons between regions of the fireball wind with large difference in
$\Gamma$ \cite{MnR_nus}. 
If, for example, plasma expanding with very high Lorentz factor,
$\Gamma>100$, is confined to a narrow jet surrounded by a slower, 
$\Gamma\sim10$ wind, internal collisions within the slower wind can heat
neutrons to relativistic temperature, leading to significant diffusion
of neutrons from the slower wind into the faster jet. Such process may
operate for winds with $\eta<400$ as well as for $\eta>400$, and may
lead, for certain (reasonable) wind parameter values, to $\sim10$~GeV
neutrino flux similar to that due to $p$-$n$ decoupling in a 
$\eta>400$ wind.

\subsection{Implications}
\label{sec:implications}

The high energy neutrinos predicted in the dissipative wind model of GRBs
may be observed by detecting the Cerenkov light emitted by high energy muons
produced by neutrino interactions below a detector on the surface of the
Earth (see \cite{GHS95} for a recent review). 
The probability $P_{\nu\mu}$ that a neutrino
would produce a high energy muon in the detector is approximately given by 
the ratio of the high energy muon range to the neutrino mean free path.
For the neutrinos produced in internal shocks, $\epsilon_\nu\sim10^{14}$~eV, 
$P_{\nu\mu}\simeq1.3\times10^{-6}(\epsilon_\nu/1{\rm TeV})$ \cite{GHS95}.
Using (\ref{eq:JGRB}), the expected flux of neutrino induced muons is
\begin{equation}
J_{\mu}\approx10{f_\pi(\epsilon_{pb})\over0.2}
{\rm km}^{-2}{\rm yr}^{-1}\,.
\label{eq:Jmu}
\end{equation}
The rate is almost independent of $\epsilon_{\nu b}$, due to the increase 
of $P_{\nu\mu}$ with energy. 
The rate (\ref{eq:Jmu}) is comparable to
the background expected due to atmospheric neutrinos \cite{GHS95}.
However, neutrino bursts should be easily detected above the background, 
since the
neutrinos would be correlated, both in time and angle, with the GRB
$\gamma$-rays. A ${\rm km}^2$ neutrino detector should detect each year
$\sim10$ neutrinos correlated with GRBs. Note, that at the high energies
considered, knowledge of burst direction and time will allow to discriminate
the neutrino signal from the background by looking not only
for upward moving neutrino induced muons, but also by looking
for down-going muons.

The predicted flux of $\sim10^{17}$~eV neutrinos, produced
by photo-meson interactions during the onset of fireball interaction 
with its surrounding medium, Eqs. (\ref{eq:JGRBAG},\ref{eq:JGRBAGw}), 
may be more difficult to detect. 
For the energy range of afterglow neutrinos, 
the probability $P_{\nu\mu}$ that a neutrino
would produce a high energy muon with the currently
required long path within the detector is
$P_{\nu\mu}\approx3\times10^{-3}(\epsilon_\nu/10^{17}{\rm eV})^{1/2}$
{\cite{GHS95,Gandhi98}. This implies, 
using (\ref{eq:JGRBAG}), an expected detection rate of muon neutrinos
of $\sim0.06/{\rm km}^2{\rm yr}$ (over $2\pi$~sr), assuming fireballs
explode in and expand into typical inter-stellar medium gas.
If, on the other hand, most GRB progenitors are massive stars and fireballs
expand into a pre-existing stellar wind, Eq.
(\ref{eq:JGRBAGw}) implies a detection of
several muon induced neutrinos per year in a $1{\rm km}^3$ detector.
We note, that
GRB neutrino detection rates may be significantly higher than derived based
on the above simple arguments, because the knowledge
of neutrino direction and arrival time may relax the requirement for long
muon path within the detector.

Air-showers could be used to detect ultra-high energy neutrinos.
The neutrino acceptance of the planned Auger detector,
$\sim10^4{\rm km^3}{\rm sr}$ \cite{Auger-nus}, seems too low.
The effective area of proposed space detectors \cite{OWL1,OWL2}
may exceed $\sim10^6{\rm km^2}$
at $\epsilon_\nu>2\times10^{19}$~eV, detecting several tens of
GRB correlated events per year, provided that the neutrino flux extends to
$\epsilon_\nu>2\times10^{19}$~eV. Since, however, the GRB neutrino flux
is not expected to extend well above  $\epsilon_\nu\sim10^{19}$~eV, and since
the acceptance of space detectors decrease rapidly below
$\sim10^{19}$~eV, the detection rate of space detectors would depend
sensitively on their low energy threshold.

Detection of high energy neutrinos
will test the shock acceleration mechanism and the suggestion that
GRBs are the sources of ultra-high energy protons, since $\ge10^{14}$~eV
($\ge10^{18}$~eV)
neutrino production requires protons of energy $\ge10^{16}$~eV
($\ge10^{19}$~eV). The dependence of 
$\sim10^{17}$~eV neutrino flux
on fireball environment imply that the detection of high energy
neutrinos will also provide constraints on GRB progenitors.

Inelastic $p$-$n$ collisions may produce $\sim10$~GeV neutrinos
with a fluence of $\sim10^{-4}{\rm cm}^{-2}$ per burst, 
due to either $p$-$n$ decoupling in a wind 
with high neutron fraction and high, $>400$, 
Lorentz factor \cite{Derishev99,BnM00}, 
or to neutron diffusion in a wind with, e.g., 
strong deviation from spherical symmetry \cite{MnR_nus}.
The predicted number of events in a $1{\rm km}^3$ neutrino telescope
is $\sim10{\rm yr}^{-1}$. Such events may be detectable in a suitably
densely spaced detector. Detection of $\sim10$~GeV neutrinos will 
constrain the fireball neutron fraction, and hence the GRB progenitor.

Detection of neutrinos from GRBs could be used to
test the simultaneity of
neutrino and photon arrival to an accuracy of $\sim1{\rm\ s}$
($\sim1{\rm\ ms}$ for short bursts), checking the assumption of 
special relativity
that photons and neutrinos have the same limiting speed.
These observations would also test the weak
equivalence principle, according to which photons and neutrinos should
suffer the same time delay as they pass through a gravitational potential.
With $1{\rm\ s}$ accuracy, a burst at $100{\rm\ Mpc}$ would reveal
a fractional difference in limiting speed 
of $10^{-16}$, and a fractional difference in gravitational time delay 
of order $10^{-6}$ (considering the Galactic potential alone).
Previous applications of these ideas to supernova 1987A 
(see \cite{John_book} for review), where simultaneity could be checked
only to an accuracy of order several hours, yielded much weaker upper
limits: of order $10^{-8}$ and $10^{-2}$ for fractional differences in the 
limiting speed and time delay respectively.

The model discussed above predicts the production of high energy
muon and electron neutrinos. 
However, if the atmospheric neutrino anomaly has the explanation it is
usually given, oscillation to $\nu_\tau$'s with mass $\sim0.1{\rm\ eV}$
\cite{Casper91,Fukuda94,Fogli95}, then
one should detect equal numbers of $\nu_\mu$'s and $\nu_\tau$'s. 
Up-going $\tau$'s, rather than $\mu$'s, would be a
distinctive signature of such oscillations. 
Since $\nu_\tau$'s are not expected to be produced in the fireball, looking
for $\tau$'s would be an ``appearance experiment.''
To allow flavor change, the difference in squared neutrino masses, 
$\Delta m^2$, should exceed a minimum value
proportional to the ratio of source
distance and neutrino energy \cite{John_book}. A burst at $100{\rm\ Mpc}$ 
producing $10^{14}{\rm eV}$ neutrinos can test for $\Delta m^2\ge10^{-16}
{\rm eV}^2$, 5 orders of magnitude more sensitive than solar neutrinos.

\paragraph*{Acknowledgments.} 

This work was supported in part by grants
from the Israel-US BSF (BSF-9800343), MINERVA, and AEC (AEC-38/99).
EW is the Incumbent of the Beracha
foundation career development chair.

\end{document}